\documentstyle[aps]{revtex}

{\end{list}}
\newcounter{enumct}

{\begin{list}{}{\setlength{\topsep}{0mm} \setlength{\itemsep}{0mm}
\setlength{\parskip}{0mm} \setlength{\parsep}{0mm}
\setlength{\leftmargin}{20mm} \setlength{\rightmargin}{0mm}
\setlength{\labelwidth}{18mm} \setlength{\labelsep}{2mm}}}%
{\end{list}}
{\begin{list}{}{\setlength{\topsep}{0mm} \setlength{\itemsep}{0mm}
\setlength{\parskip}{0mm} \setlength{\parsep}{0mm}
\setlength{\leftmargin}{10mm} \setlength{\rightmargin}{0mm}
\setlength{\labelwidth}{18mm} \setlength{\labelsep}{2mm}}}%
{\end{list}}

\newlength{\captivewidth}
\newlength{\tablinsep}
\newlength{\halfpagewid}
\input{epsf}

\begin{document}

\sloppy
\setcounter{page}{0}
\thispagestyle{empty}

\begin{flushright}
nucl-th/9705041
\end{flushright}
\vspace{3.5cm}

\begin{center}
{\Large
{\bf INTERPLAY OF PARTON AND HADRON CASCADES
\\${}$\\
IN NUCLEUS-NUCLEUS COLLISIONS 
\\${}$\\
AT THE {\sl CERN SPS} AND {\sl RHIC}}
}
\end{center}
\bigskip

\begin{center}

{\Large
{\bf  Klaus Geiger and Ronald  Longacre}
}
\medskip

{\it Physics Department,
Brookhaven National Laboratory, Upton, N.Y. 11973, U.S.A.}
\end{center}
\smallskip

\begin{center}
e-mail: klaus@bnl.gov, longacre@bnl.gov  \\
\end{center}
\vspace{0.5cm}

\begin{center}
{\large {\bf Abstract}}
\end{center}
\bigskip

We introduce a Monte Carlo space-time model for high-energy collisions 
with nuclei, involving the dynamical interplay of perturbative QCD  
parton production and evolution, with non-perturbative  parton-cluster 
formation and `afterburner' cascading of formed pre-hadronic clusters 
plus hadron excitations.  This approach allows us to trace the space-time 
history of parton and hadron degrees of freedom of nuclear collisions 
on the microscopical level of parton and hadron cascades in both 
position and momentum space, from the instant of nuclear overlap to the 
final yield of particles.  In applying this approach, we  analyze $Pb+Pb$ 
collisions at the CERN SPS with beam energy 158 GeV ($\sqrt{s}/A = 17$ GeV)
and $Au+Au$ collisions at RHIC with  collider energy $\sqrt{s}/A = 200$ GeV.
We find that the perturbative QCD parton production and cascade development
provides an important contribution to particle production at central 
rapidities, and that the `afterburner' cascading of pre-hadronic clusters 
and formed hadrons emerging from the parton cascade is  essential.
The overall agreement of our model calculations including the `afterburner' 
cascading with the observed particle spectra at the CERN SPS is fairly good, 
whereas the neglect of the final-state interactions among hadronic 
excitations deviates significantly.
\noindent
\bigskip

\newpage

\section{INTRODUCTION}
\label{sec:section1}
\bigskip

The physics of QCD at high density has become in the last years a most
important field of research, for example, in the context of
the rise of the gluon structure functions in deeply inelastic $ep$ scattering 
at very small $x$ \cite{hera},
the propagation of hard probes in dense nuclear matter such as jets \cite{wanghuang}
 or charmonium \cite{kharzeev}, or, the issue of formation and evolution
of a quark-gluon plasma in collisions of heavy nuclei \cite{bm}.
Hence, high-density QCD effects are notoriously  present,
 in one way or another, in high-energy collision experiments 
that involve interactions of hadronic or nuclear matter,
hence offering a diverse resort of exciting new physics.

Specifically the experimental heavy-ion programs at the CERN SPS (currently under way),
the RHIC and LHC colliders (to be launched in the next few years), are concepted
in the hope of discovering novel collective behavior of highly compressed 
nuclear matter through characteristic particle production from this dense matter.
In this scope, the
prime motivation for ultra-relativistic heavy-ion collisions is
to create a state of high particle- and energy-density, and to study the decay of the
system to gain knowledge of QCD interactions under these extreme conditions.
The characteristics of particle production in
the environment of dense, strongly interacting matter should 
teach us about significance and properties of the interplay
between partonic and hadronic degrees of freedom,
i.e., fundamental questions concerning
deconfinement dynamics, formation and evolution
of a quark-gluon plasma, the nature of the QCD phase-transition between
quark-gluon and hadron phases, et cetera.
\smallskip

Clearly, to extract unambigous signatures of 
these new physics aspects from `messy' nuclear collision events with
several hundreds, or even thousands  of final-state particles,
a good understanding of the microscopic dynamics is required,
that is, the evolution from the initial state of colliding nuclei, via
parton and hadron production, up to the final appearance of particles
in the detectors.
Here lies the essential concern of
this paper, namely,
to study the role and relative importance of partons and  hadrons, as well
as their dynamical interplay,
in a generally mixed parton-hadron system 
created by a high-energy collision of two heavy nuclei.
To do so,
we take  steps towards a  phenomenological,  QCD-based,
description that combines  quark-gluon {\it and} hadron degrees of freedom
in the evolution of nucleus-nucleus collisions
at CERN SPS energies and above.
\smallskip

To the best of our knowledge  no constructive attempt has been made before
to merge the perturbative QCD parton-cascade picture with a hadron-cascade model.
On the contrary, the  event generators that are currently on the market
for desribing nucleus-nucleus collisions are exclusivily based
on  modelling  nuclear
collisions in terms of nucleon-nucleon collisions on the
basis of a constituent valence-quark picture plus
string-excitation and -fragmentation.
Popular examples for these models are
FRITIOF \cite{fritiof}, 
VENUS \cite{venus}, 
RQMD \cite{rqmd},
DPM \cite{dpm}. 
Distinct from these is  HIJING \cite{hijing}, 
which also incorporates a perturbative QCD approach to multiple minijet
production, however, it does not incorporate a space-time description.
In particular VENUS and RQMD are  impressively fine-tuned to describe the recent
CERN SPS data. Yet, the price paid is the invention
of supplemental mechansims (such as `string droplets' in VENUS or `color ropes' in RQMD)
to mimic certain  underlying dynamics due to nuclear effects, which
cannot be accounted for in non-trivial manner.
\smallskip

Our rationale is different: We strictly refrain from
introducing  ad-hoc recipes to parametrize physics details that may be
novel in nuclear collisions as compared to hadronic collisions.
Rather, we wish to examine in how far
a well-defined semiclassical  multi-particle description in terms of
partons and hadrons suffices to accomodate
the bulk characteristics of the CERN data, and hence
may serve as a prospect for   RHIC and LHC heavy-ion physics.
We will not address here the fundamental nature of complex underlying
physics aspects; instead we adopt a probabilistic particle
description with dynamically changing proportions of partons and hadrons, and develop
a space-time model that allows to simulate nucleus collisions
from the first instant of overlap, via QCD parton cascade development
at the early stage,
parton conversion into pre-hadronic excitations and hadron
formations, up to the cascading of these pre-hadrons and final state
particles, as well as the beam remnant nucleons at late times. 
As we shall show, this approach does remarkably well in comparison
to the gross particle production properties observed at CERN SPS.
No attempt is made to fine-tune our model to the data, and hence
we presume that a more detailed analysis than carried out in this paper,
may exhibit  disagreement, in which case
this may be taken as indicator of truly new physics.
\smallskip

To set the stage, we recall  the standard picture of
high-energy hadron collisions as compared to hadron-nucleus
or nucleus-nucleus collisions
(for convenience we view  the collisions in the overall center-of-mass frame): 
\begin{description}
\item{$\bullet$}
In {\it hadron-hadron collisions} with large momentum transfer,
the coherence of the incoming hadron wavefunctions
is (at least partially) destroyed upon collisional contact
by means of one or more hard parton scatterings and/or 
by soft non-diffractive interactions between the beam hadrons.
The liberated partons that materialize mostly around midrapidity
after a hard scattering, evolve as jets (accompanied
by gluon radiation), and eventually reorganize 
to color-singlet systems that form
final state hadrons around central rapidities $|y|\approx y_{cent} \equiv 0$.
On the other hand, the 
remnants of the initial hadrons that have not taken
part in the hard process, recede down the beam pipe with
little energy loss, and produce hadron fragments around
$|y| \, \lower3pt\hbox{$\buildrel <\over\sim$}\,|y_{beam}|$,
near the beam rapidities $ y_{beam} = \pm \ln(\sqrt{s}/M_h)$.
The dynamics of parton evolution and hadron production at small
rapidities should therefore decouple to good approximation
from the fragmentation of the beam remnants at large
rapidities.
As is well known,
such a behaviour is indeed observed at $p\bar{p}$ collider experiments
with
$\sqrt{s}  \, \lower3pt\hbox{$\buildrel >\over\sim$}\,100$ GeV,
in which the measured baryon rapidity distributions shows
two seperate peaks around $\pm y \approx y_{beam}$ with a 
distinct gap in between,
whereas the rapidity distributions of mesons is strongly
concentrated in the central region.
This characteristic behaviour of hadron-hadron
collisions is usually termed {\it transparency}.
\item{$\bullet$}
In {\it nucleus-nucleus collisions} at 
beam energies per nucleon comparable to proton-proton
colliders, the picture of a fairly clean separation of the three
rapidity regions (central, forward and backward beam fragmentation regions)
 as in hadron collisions breaks down.
Nonetheless, the basic dynamics of particle production 
parallels  those in hadronic collisions, except that it is amplified 
by the large interaction volume of the nuclear collision system.
Hence, much larger number of hard parton
scatterings is likely to be initiated by sufficiently large momentum transfer
collisions of nucleon pairs from
the two beam nuclei, leading to copious parton materialization and
(mini)jet production in the central rapidity region.
The beam remnants containing the spectator nucleons 
on the other hand populate the forward and backward
rapidity regions, but to lesser extent than in hadronic collisions
at comparable energy.
However, it is a well known experimental fact that
in nucleus-nucleus collisions a rigid division
into the three rapidity regions is not possible anymore.
Rather than transparency, one observes a significant {\it stopping}
associated with slowing down of the initial nucleons
due to inelastic nucleon-nucleon collisions,
on top of the nucleons energy loss from hard parton production.
Moreover, there are effects of {\it rescattering} among
partons, among initial state nucleons, as well as among
newly produced final state hadrons. 
Rescattering  consequently must lead
to  an enhanced particle production and $p_\perp$-broadening 
in the central region, whereas stopping
causes a significant baryon population at midrapidity
as well as a smearing of the peaks at forward and
backward beam rapidities.
\end{description}
\medskip

The preceding qualitative discussion serves us as motivation
for investigating the space-time dynamics
of nucleus-nucleus collisions within a combined parton-hadron
cascade model, raising  the following two key questions:
\begin{description}
\item[1.]
To which extent undergo the colliding nuclei 
materialization of partonic constituents,
what is the relative importance of associated parton cascading,
and how does the simultanous evolution of co-existing
partonic and hadronic matter proceed in space and time?
\item[2.]
What is the cumulate effect on final-state
particle distributions from
the space-time history of
rescatterings among partons during the early stage,
and from re-interactions involving the remnant nucleons
as well as  newly formed hadronic excitations at later times?
\end{description}
\medskip

The paper is organized as follows. 
In Sec. II, we introduce in brevity some formal aspects of multi-particle
kinetic theory, which form the theoretical basis of our QCD space-time cascade model.
Sec. III is devoted to the application of the model to heavy-ion
coollisions, first to $Pb+Pb$ collsions at CERN SPS energy of $\sqrt{s}/A = 17$ GeV,
and then
to $Au+Au$ collsions at RHIC energy of $\sqrt{s}/A = 200$ GeV.
Sec. IV contains concluding remarks and discusses some future perspectives.

\bigskip
\bigskip

\section{DESCRIPTION OF THE MODEL}
\label{sec:section2}
\bigskip

In this Section we explain the main aspects of our
parton-hadron cascade model. We will not be concerned 
here with
the elaborate details of the model, but refer the interested
reader to the extensive documentations of Refs.
\cite{ms44,msrep} for  the parton cascade aspects and
Refs. \cite{hijet} for the hadron cascade features. 
The central element in our 
space-time cascade description is the use of
relativistic transport theory \cite{msrep} in conjunction
with renormalization-group  improved QCD \cite{ms39},
which provides the theoretical basis
to follow the QCD evolution 
in 7-dimensional phase-space $d^3 r d^3 k dE$
of a mixed 
multi-particle system of partons and hadrons
with dynamically changing proportions.

As we explain further below, we combine the following
four elements:
(i) the {\it initial state} of the nucleus-nucleus collision
system in terms of ncleon and parton degrees of freedom,
(ii)the {\it  parton-cascade development}
which embodies the perturbative QCD evolution of 
multiple parton collisions
including inelastic (radiative) processes,
(iii)
the phenomenological
{\it parton-hadron conversion} model of Ellis and Geiger  \cite{ms37,ms40,ms41},
in which the hadronization mechanism is described in terms of 
dynamical parton-cluster formation with subsequent  decay of 
color-singlet clusters into hadrons,
and, 
(iv)
the {\it `afterburner' hadron cascade} of produced pre-hadronic clusters and hadrons
which incorporates cluster-cluster, cluster-hadron, hadron-hadron collisions,
as well as resonance formation and decay.

The microscopic history of the dynamically-evolving particle system
is traced in space-time {\it and} momentum space, so that
the correlations of particles in space,  time, color and flavor can be taken
into account systematically.
We emphasize that the interplay
between perturbative and non-perturbative regimes is controlled locally
by the space-time evolution of the mixed parton-cluster-hadron system itself
(i.e., the time-dependent local particle densities),
rather than by an arbitrary global division
between parton and hadron degrees of freedom 
(i.e., a parametric energy/momentum cut-off).
In particular the parallel evolution of the mixed system
of partons, pre-hadronic clusters, and hadrons, with the relative proportions
determined by the dynamics itself, is a novel feature that
is only possible by keeping track of both space-time and 
energy-momentum variables.

The model as a whole consists of
four major building-blocks, describing the above-mentioned evolution stages of
a nucleus-nucleus collision
from the initial  beam/target collision system
upon collisional  contact, through the QCD-evolution of
parton distributions, hadron formation and cascading,
up to the emergence of final hadronic states:
\begin{description}
\item[(i)]
The {\it initial state} associated with the incoming
nuclei involves their decomposition into nucleons and of the nucleons into
partons on the basis of the experimentally measured  
nucleon structure functions and elastic form-factors.
This procedure translates the initial nucleus-nucleus system into
two colliding clouds of {\it virtual} partons according
to the well-established parton decpomposition of
the nuclear wavefunctions at high energy \cite{GLR}.
\item[(ii)]
The {\it parton cascade development}
starts from the initial interpenetrating parton clouds,
and involving  the space-time
development with mutual- and self-interactions of the 
system of quarks and gluons.
Included are multiple elastic and inelastic interaction processes, described 
as sequences of elementary $2 \rightarrow 2$ scatterings, $1\rightarrow 2$
emissions and $2 \rightarrow 1$ fusions.
Moreover,  correlations are accounted for between primary virtual
partons, emerging as unscathed remainders from the initial state, and
secondary real partons, materialized or  produced 
through the partonic interactions.
\item[(iii)]
The {\it hadronization dynamics} of the evolving system
in terms of parton-coalescence to color-neutral clusters
is decribed as a local, statistical process that depends on the spatial separation
and color of nearest-neighbor partons.
Each pre-hadronic parton-cluster fragments through isotropic two-body decay
into  primary hadrons, according to the density of
states, followed by the decay of the latter into final 
stable hadrons.
\item[(iv)]
The {\it `afterburner' hadron cascade} describes the evolution
of produced pre-hadronic clusters
and hadrons, emerging both from the hadronization
of cascading partons, as well as from primary remnant partons which
represent the fraction of unscathed initial state nucleons.
Pre-hadronic clusters can mutually rescatter, or scatter off close-by
hadrons, before they decay into stable final-state hadrons.
Similarly, already formed hadrons may deflected by
elestic collisions with other hadrons and clusters, or may be excited by inelastic
collisions and resonance formation/decay.
\end{description}
\medskip

A firm theoretical basis for
the above space-time cascade description of a multiparticle system
in high-energy collisions can be derived systematically from
{\it quantum-kinetic theory} on the basis of
QCD's first principles in a stepwise approximation scheme 
(see  e.g., Refs. \cite{ms39,ms42} and references therein).
This framework allows  to
cast the time evolution of the mixed system of
individual partons, composite parton-clusters, and physical hadrons
in terms of a closed set of
integro-differential equations for
the phase-space densities of the different particle excitations.
The definition of these phase-space densities,
denoted by
$F_\alpha$, where $\alpha\equiv p, c, h$
labels the species of partons, pre-hadronic clusters, or hadrons,
respectively, is:
\begin{equation}
F_\alpha(r,k)\;\,\equiv\; \, F_\alpha (t, \vec r; E, \vec k)
\;\,=\;\,
\frac{dN_\alpha (t)}{d^3r d^3k dE}
\;,
\label{F}
\end{equation}
where $k^2 = E^2 -\vec{k}^{\,2}$ can be off-shell 
(space-like $k^2 < m^2$, time-like $k^2 > m^2$) or on-shell ($k^2 = m^2$).
The densities (\ref{F}) measure the number of particles
of type $\alpha$ at time $t$ with position in $\vec r + d\vec{r}$,
momentum in $\vec k + d\vec{k}$,
and energy in $E + dE$ (or equivalently invariant mass in $k^2 + dk^2$).
The $F_\alpha$ are the quantum analogues of the
classical phase-space distributions, including both off-shell and on-shell
particles, and hence
contain the essential microscopic
information required for a statistical description
of the time evolution of a many-particle system in
complete 7-dimensional phase-space $d^3rd^3kdE$, 
thereby providing the basis for calculating
macroscopic observables.

The phase-space densities (\ref{F}) are determined by the
self-consistent solutions of
a set of {\it transport equations} (in space-time) coupled with
renormalization-group type {\it evolution equations} (in momentum space).
Referring  to Refs. \cite{ms37,ms39} for details,
these equations can be generically expressed as
convolutions of the densities $F_\alpha$ of particle species $\alpha$,
interacting with  specific cross sections $\hat{I}_j$ for the processes $j$.
The resulting coupled equations for the
space-time development of
the densities of partons $F_{p}$, clusters $F_c$ and
hadrons $F_h$ is a self-consistent set in which the change
of the densities $F_\alpha$ is governed by the balance of
the various possible interaction processes among the particles.
As compared to Ref. \cite{ms44}, here we consider
in novel addition  the `afterburner' cascade of clusters and hadrons as mentioned before.
Figs. 1-3 represents these equations pictorially.
For the densities of  partons $F_p$,
the {\it transport equation} 
(governing the space-time change with $r^\mu$) and the {\it evolution equation} 
(controlling the change with momentum scale $k^\mu$), read, respectively, (Fig. 1),
\begin{eqnarray}
k_\mu \frac{\partial}{\partial r^\mu}\; F_p(r,k)
&=&
F_{p''} F_{p'''}\circ 
\left[\frac{}{}
\hat{I}(p''p'''\rightarrow p p') \;+\;\hat{I}(p''p'''\rightarrow p)
\right]
\;-\; 
F_{p} F_{p'}\circ 
\left[\frac{}{}
\hat{I}(pp'\rightarrow p'' p''')\;+\; \hat{I}(pp'\rightarrow p'')
\right]
\nonumber \\
& &
-\;\;
F_{p} F_{p'}\circ \left[\frac{}{} 
\hat{I}(p'p''\rightarrow p)\;-\; \hat{I}(pp'\rightarrow p'')\right]
\;-\;
F_p\,F_{p'}\circ \hat{I}(p p'\rightarrow c)
\label{e1}
\\
k^2  \frac{\partial}{\partial k^2}\; F_p(r,k)
&=&
F_{p'}\circ \hat{I}(p'\rightarrow p p'')\;-\;
F_{p}\circ \hat{I}(p\rightarrow p' p'')
\label{e2}
\;.
\end{eqnarray}
For the densities of {\it pre-hadronic clusters}  and
{\it hadrons}, the evolution equations  are homogeneous
to good approximation,
so that one is left with non-trivial transport equations only,
\footnote{
It is worth noting that eq. (\ref{e2})  embodies the momentum space
($k^2$) evolution of partons through
the renormalization of the phase-space densities $F_p$, determined
by their change $k^2 \partial F_p(r,k)/\partial k^2$
with respect to a variation of the mass (virtuality) scale $k^2$
in the usual QCD evolution framework \cite{dok80,jetcalc,bassetto}.
On the other hand,
for pre-hadronic clusters and hadrons, renormalization effects
are comparatively small, so that their
mass fluctuations $\Delta k^2/k^2$ can be ignored to first
approximation,
implying $k^2 \partial F_c(r,k) /\partial k^2
= k^2 \partial F_h(r,k) / \partial k^2  =0 $.
}.
For the evolution of the cluster densities $F_c$, we have (Fig. 2)
\begin{eqnarray}
k_\mu \frac{\partial}{\partial r^\mu}\; F_c(r,k)
&=&
F_p\,F_{p'}\circ \hat{I}(p p'\rightarrow c)
\;-\;
F_c\circ \hat{I}(c\rightarrow h)
\;+\;
F_{c''} F_{c'''}\circ \hat{I}(c''c'''\rightarrow c c') \;-\; 
F_{c} F_{c'}\circ \hat{I}(cc'\rightarrow c'' c''')
\label{e3a}
\\
& &
\;+\; 
F_{c'} F_{h'}\circ \hat{I}(c'h'\rightarrow c h) \;-\; 
F_{c} F_{h}\circ \hat{I}(ch\rightarrow c' h')
\;\;+\;\;\ldots\ldots
\nonumber
\\
k^2  \frac{\partial}{\partial k^2}\; F_c(r,k) &=& 0  
\label{e3b}
\;,
\end{eqnarray}
and similarly, for the evolution of the hadron densities $F_h$, the equations read
(Fig. 3)
\begin{eqnarray}
k_\mu \frac{\partial}{\partial r^\mu}\; F_h(r,k)
&=&
F_c \circ\hat{I}(c\rightarrow h)
\;+\;
\left[\frac{}{}
F_{h'}\circ \hat{I}(h'\rightarrow h)
\;-\;
F_h\circ \hat{I}(h\rightarrow h')
\right]
\;+\;
F_{h''} F_{h'''}\circ \hat{I}(h''h'''\rightarrow h h')
\label{e4a} \\
& &
 \;-\; 
F_{h} F_{h'}\circ \hat{I}(hh'\rightarrow h'' h''')
\;+\; 
F_{c'} F_{h'}\circ \hat{I}(c'h'\rightarrow c h) \;-\; 
F_{c} F_{h}\circ \hat{I}(ch\rightarrow c' h')
\;\;+\;\;\ldots\ldots
\nonumber
\\
k^2  \frac{\partial}{\partial k^2}\; F_h(r,k) &=& 0
\label{e4b}
\;.
\end{eqnarray}
In (\ref{e1})-(\ref{e4b}),
each convolution $F \circ\hat{I}$ of
the density of particles $F$ entering a particular vertex
${\hat I}$ includes a sum over contributing
subprocesses, and a phase-space integration
weighted with the associated subprocess probability distribution
of the squared amplitude. Explicit expressions are given in
Refs.  \cite{msrep,ms37}.
\bigskip
\bigskip

\noindent
The terms on the right-hand side
of the transport-  and evolution-equations (\ref{e1})-(\ref{e4b})
corresponds to one of the following categories (c.f. Figs. 1-3):
\begin{itemize}
\item
parton scattering and parton fusion  through 2-body collisions,
\item
parton multiplication through radiative emission processes
on the perturbative level,
\item
colorless cluster formation through parton coalescence
depending on the local color and spatial configuration,
\item
hadron formation by decay of the cluster excitations into final-state hadrons.
\item
scattering of pre-hadronic clusters or already formed hadrons
 with other clusters or  hadrons.
Note: this includes also (as indicated by the `dots') 
cluster-cluster fusion, cluster-hadron absorption,
as well as inelastic collisions among clusters and hadrons, in which
energy-momentum is transferred into excitation.
\end{itemize}
\bigskip
\bigskip

The equations (\ref{e1})-(\ref{e4b}) reflect a {\it probabilistic
interpretation} of the multi-particle evolution in 
space-time and momentum space
in terms of sequentially-ordered interaction processes $j$,
in which the rate of change of the particle distributions $F_\alpha$
($\alpha=p,c,h$)
in a phase-space element $d^3rd^4k$
is governed by the balance of gain (+) and loss ($-$) terms.
The left-hand side
describes free propagation of a
quantum of species $\alpha$, whereas
on the right-hand side the interaction kernels $\hat{I}$
are integral operators that incorporate the effects of
the particles' self-  and mutual interactions.
This probabilistic  character 
is essentially an effect of time dilation, because in any frame
where the particles move close to the speed of light, the associated
wave-packets are highly localized to short space-time extent, so that
comparatively
long-distance quantum interference effects are generally small.

\newpage 

${}$
\vspace{2.0cm}
\begin{figure}
\epsfxsize=450pt
\centerline{ \epsfbox{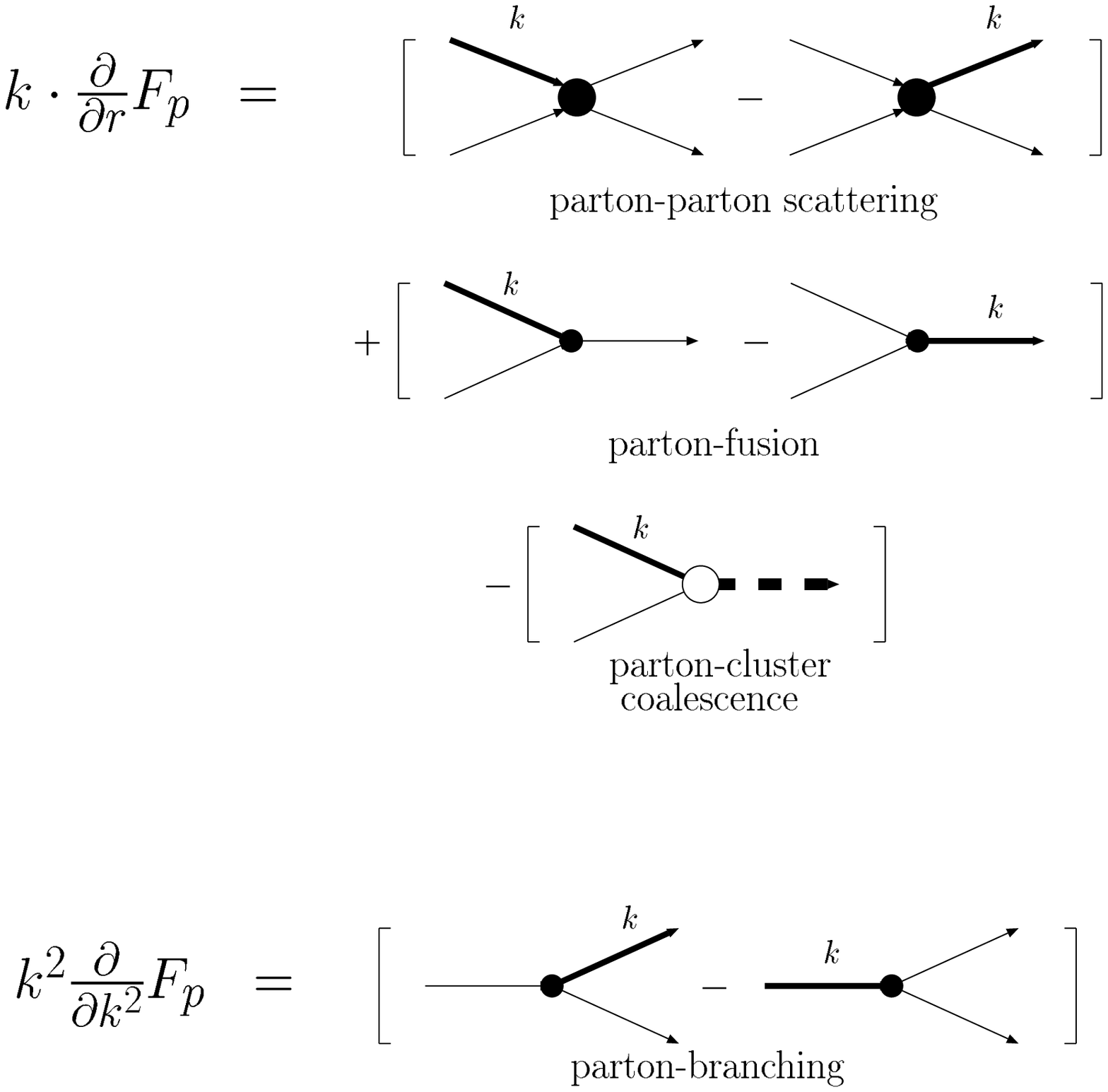} }
\vspace{-4.0cm}
\caption{
Graphical representation of the equations (2) and (3) for the
particle phase-space densities $F_p$ of partons.
\label{fig:fig1}
}
\end{figure}

\newpage

${}$
\vspace{1.0cm}
\begin{figure}
\epsfxsize=450pt
\centerline{ \epsfbox{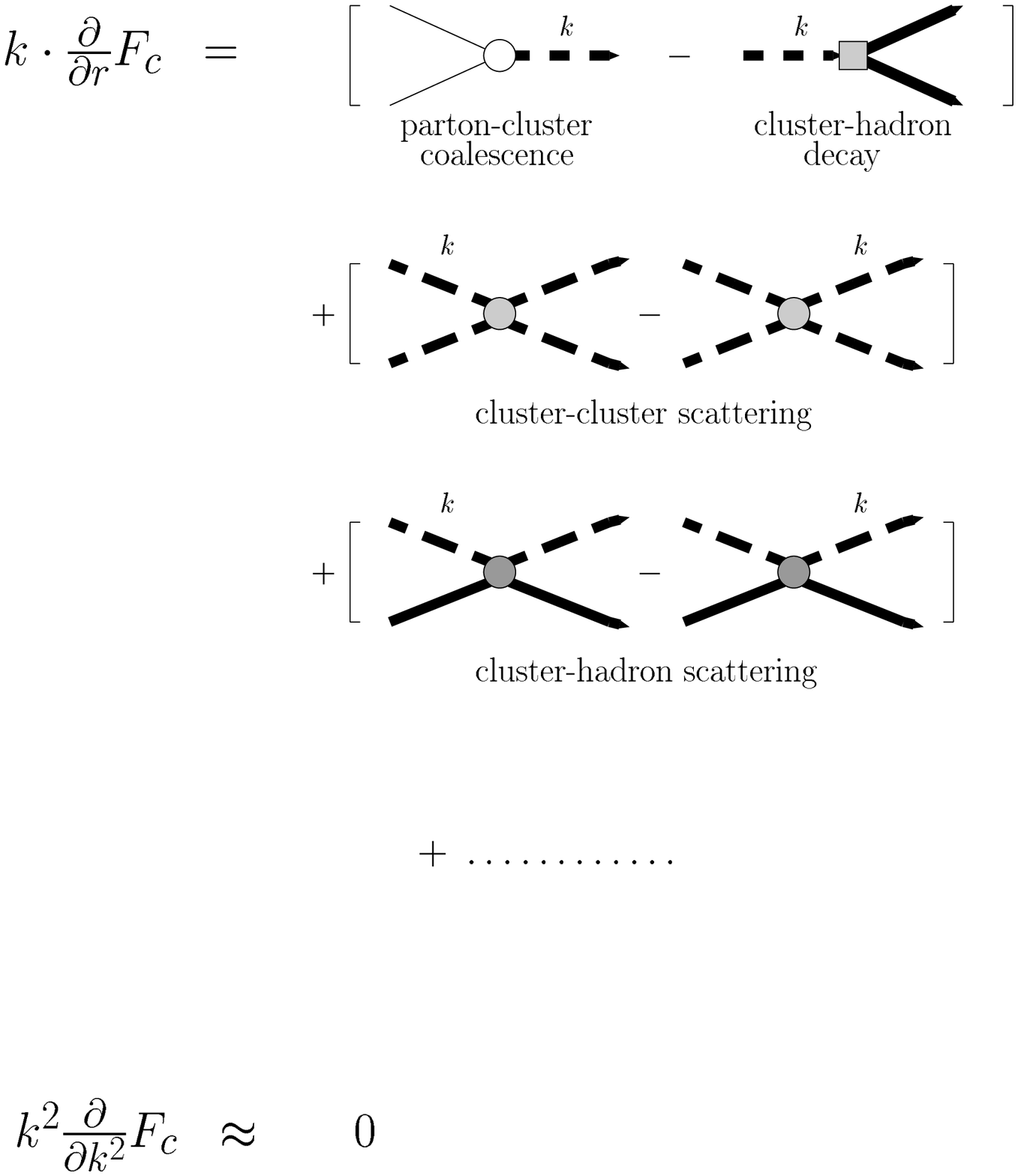} }
\vspace{-1.0cm}
\caption{
Graphical representation of the equations (4) and (5) for the
particle phase-space densities $F_c$ of pre-hadronic clusters.
The `dots' represent further processes not explicitely shown, such as
cluster-cluster fusion, cluster-hadron absorption,
as well as inelastic collisions among clusters and hadrons, in which
energy-momentum is transferred into excitation.
\label{fig:fig2}
}
\end{figure}

\newpage

${}$
\vspace{0.0cm}
\begin{figure}
\epsfxsize=450pt
\centerline{ \epsfbox{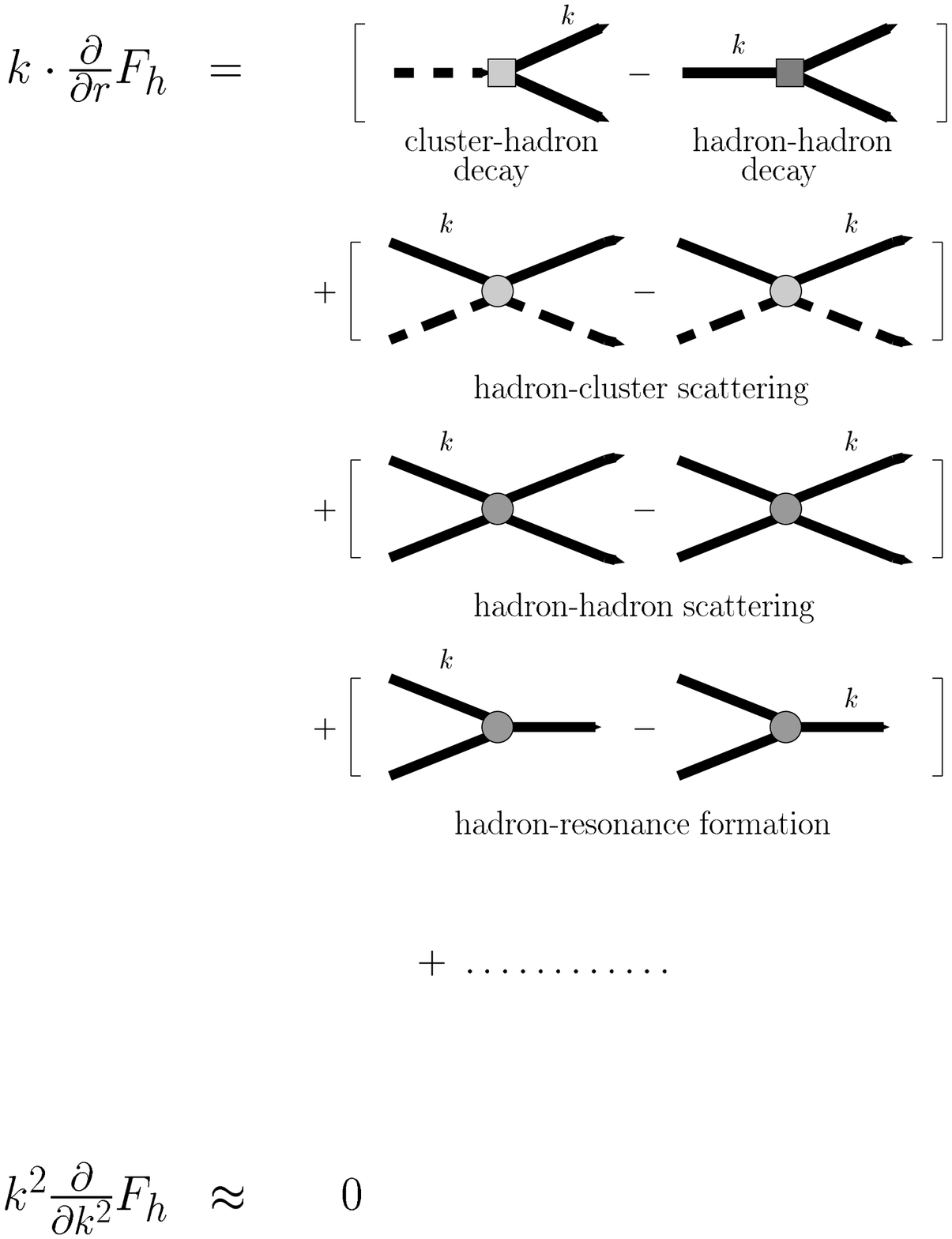} }
\vspace{0.0cm}
\caption{
Graphical representation of the equations (6) and (7) for the
particle phase-space densities $F_h$ of hadrons.
The `dots' represent further processes not explicitely shown, such as
cluster-hadron absorption,
as well as inelastic collisions among clusters  and hadrons.
\label{fig:fig3}
}
\end{figure}
\bigskip
\bigskip

\newpage 

\section{APPLICATION TO HEAVY-ION COLLISIONS AT THE 
{\sl CERN SPS} AND {\sl RHIC}}
\label{section3}
\bigskip

In this Section we apply the our approach to heavy-ion collisions,
first, to $Pb+Pb$ collisions at the nominal CERN SPS beam energy
158 GeV, corresponding to a center-of-mass energy 
per nucleon of $\sqrt{s}/A = 17$  GeV,
and second,
to $Au+Au$ collisions at the anticipated RHIC energy
per nucleon of $\sqrt{s}/A = 200$  GeV.
We restict ourselves to these
collision systems, because $Pb+Pb$ at the CERN SPS 
is the largest nuclear system at the highest
present energy that has actually been studied in experiment
with the recent data from the '96 and '97
runs hinting first evidence of new physics apsects in high-energy
nuclear collisions,
and because $Au+Au$ at RHIC is the comparable
collision system at an order of magnitude larger beam energy.
However, we emphasize that our model applys in general to any
collision system $A+B$ including proton-nucleus collisions at 
energies $\sqrt{s}/A   \, \lower3pt\hbox{$\buildrel >\over\sim$}\,10$ GeV.
In particular, we believe its strength lies in addressing
nuclear collisions at even higher energies, such as 
anticipated at the  LHC.
\smallskip

In applying our model in practice, we note that
the probabilistic character of the transport- and evolution equations
(\ref{e1})-(\ref{e4b})
allows one to solve for the phase-space densities $F_\alpha(r,k)$ by
simulating
the dynamical development as a Markovian process causally in time.
The computer simulations, the details and results of which we will discuss below,
were performed by combining
the parton- cascade/cluster-hadronization code VNI \cite{ms44}
the hadron cascade code HIJET \cite{hijet}.
Starting from the initial parton clouds 
$F_p(t_0,\vec{r},k)$,
corresponding to
the incoming nuclei upon collisional contact at time $t=t_0$,
the set of equations (\ref{e1})-(\ref{e4b}) can  be solved
in terms of the evolution of the phase-space densities $F_p$, $F_c$ and $F_h$
of partons, clusters, and hadrons, respectively,
for $t > t_0$. 
For calculational convenience, it is most suitable
to choose the {\it  center-of-mass ($cm$) frame of the colliding 
beam particles}, with the collision axis in the $z$-direction.
We remark that in general the space-time description of particle evolution is
Lorentz-frame dependent, however, since we are concerned in the following
only with Lorentz-invariant final-state hadron spectra in rapidity and transverse
mass, the particular choice of frame is irrelevant for the results and just a
matter of preference.
\bigskip

\subsection{The simulation procedure}
\smallskip

The simulation of the time development of the mixed system
of partons, clusters, and hadrons
in position and momentum space on the basis of 
eqs. (\ref{e1})-(\ref{e4b})
emerges then from following each individual particle through its history with
the various  probabilities and time scales of interactions 
sampled stochastically from the relevant probability distributions
in the kernels $\hat{I}$ of eqs. (\ref{e1})-(\ref{e4b}):
The microspcopic history of the system can thus be traced by evolving
the phase-space distributions of particles are
evolved in small time steps ($\Delta t \simeq 10^{-3}\;fm$)
 and 7-dimesional phase-space $d^3rd^3kdE$ throughout 
the stages of parton cascade, parton-cluster formation, cluster-hadron decays,
and afterburner hadron cascade,
until stable final-state hadrons and other particles (photons, leptons, etc.) 
are left as freely-streaming particles.
The essential ingredients in this Monte-Carlo procedure are summarized as follows
\cite{ms44}:
\begin{description}
\item[(i)]
The {\it initial state} is constructed in three steps. First, the nuclei
are decomposed into the nucleons with an appropriate
Fermi-distribution. Second, the nucleons are in turn decomposed into
their parton substructure according to proton/neutron structure functions
\footnote{
We use the GRV structure function parametrization \cite{grv},
which describes quite accurately the HERA data even at low $Q^2$ and
very small $x$.
}
with a spatial distribution given by the Fourier transform 
of the nucleon elastic form-factor.
Third, the so-initialized phase-space densities of
(off-shell) partons are then boosted with
the proper Lorentz factor to the center-of-mass frame of the
colliding nuclei.
\item[(ii)]
The {\it parton cascade} development
proceeds then by
propagating the partons along classical trajectories until they interact,
i.e., collide (scattering or fusion process),
decay (emission process) 
or coalesce to pre-hadronic composite
states (cluster formation). Both space-like and time-like radiative
corrections are included within the Leading-Log approximation.
The relevant interaction probabilities are obtained from the 
well-known perturbative QCD cross-sections \cite{field}, 
and
the coalescence probabilities of the Ellis-Geiger model \cite{ms37,ms40,ms41},
respectively.
Both the production of partons and the emergence of pre-hadronic clusters
through their coalescence are subject to an individually specific  formation time
$\Delta t_{p,c} = \gamma / M_{p,c}$ where $1/M_{p,c}=1/\sqrt{k^2}$ is the proper
decay time of off-shell partons or clusters with invariant mass $M_p$,
respectively  $M_c$, and $\gamma=E/M_{p,c}$ is the Lorentz factor.
\item[(iii)]
The {\it afterburner hadron cascade} evolves similarly by propagating
the  pre-hadronic parton-clusters (those  emerging from coalescence of 
materialized, interacted partons) 
along classical paths until they either scatter off or absorb
other clusters or hadrons, or, until they convert into
primary hadrons (cluster decay), followed by the hadronic decays
into stable final state particles.
The corresponding interaction probabilities are calculated 
from the `additive quark model' \cite{aqm} within which one can
associate a specific cross-section for any pair of colliding hadrons 
according to their valence quark content.
Cluster-cluster and cluster-hadron collisions are straightforward to include
in this approach, since in our model each pre-hadronic cluster that emerges from parton
coalescence has a definite quark content, depending on the flavor of the
mother partons.
Hence, collisions among clusters and hadrons are treated on equal footing.
Finally, the the decays of excited hadrons and resonances which are formed in
these collisions, are sampled from the particle data tables \cite{pada}.
Again, each newly produced hadron becomes a `real' particle only after
a characteristic formation time 
$\Delta t_{h} = \gamma / M_{h}$ depending on their invariant mass $M_h$ and
their energy through $\gamma= E/M_h$. Before that time has passed, a hadron
must be considered as a still virtual object that cannot interact
incoherently until it has formed according to the uncertainty principle.
\item[(iv)]
The {\it beam remnants}, being the unscathed remainders of the initial nuclei, 
emerge  from reassembling all those  remnant primary partons that
have been spectators without interactions throughout the evolution.
The recollection of those yields two corresponding beam clusters
with definite charge, baryon number, energy-momentum and positions as given by
the sum of their constituents.
These beam clusters  decay into final-state hadrons which recede
along the beam direction at large rapidities of the beam/target fragmentation regions.
Again rescatterings,  absorptive and emissive processes, as well as
individual formation times, of the produced hadrons are accounted for 
as described in (iii).
\end{description}
We note that the spatial density and the momentum distribution
of the particles are intimately connected: The momentum 
distribution continously changes through the interactions and
determines how the quanta propagate in coordinate space.
In turn, the probability for subsequent interactions depends on the 
resulting local particle density. Consequently, the development
of the phase-space densities is a complex
interplay, which - at a given point of time - contains implicitely the
complete preceding history of the system.
\bigskip

\subsection{CERN SPS: central $Pb+Pb$ collisions at $E_{cm}=17$ A GeV}

We first turn to the discussion of the results 
for $Pb+Pb$ collisions at beam momentum
of $E_{beam}=158$ GeV corresponding to $\sqrt{s}/A = 17$ GeV, as
obtained from the computer simulations
according to the above prescriptions. 
\medskip

\subsubsection{{\bf Significance of parton production}}
\smallskip

To begin our discussion,
we resurrect the first question that we raised in the Introduction,
namely, how much of the nuclear collision energy is harnessed in truly
partonic materialization with subsequent parton cascading,
and how does the entwined evolution
of partonic and hadronic matter components proceed.
\smallskip

Fig. 4 sheds light on the first part of the question,
by exhibiting the relative importance
of the truly partonic contribution to the final-state
hadron yield. Shown is rapidity distribution for pion production 
$\pi^++\pi^-+\pi_0$ 
(representative for the bulk of produced hadrons) that results 
directly from parton materialization and the associated cascade evolution,
as compared to the yield of final pions that arise 
from the soft fragmentation of the nuclear beam remnants which
involves the spectator partons surviving from the initial state
without materialization. 
The `circles' give the total amount of produced pions, i.e. the
sum of the contributions from
the hadronization of materialized cascading partons via cluster
formation and decay, plus the fragmentation of the nuclear beam remnants 
involving the remaining fraction of non-materialized spectator partons.
The `squares', on the other hand, represent the contribution
that results solely from the latter beam fragmentation.
The difference between the `square' and `circle' histograms therefore reflects the
significance of the truly partonic dynamics of parton production and evolution.
The essence of Fig. 4 is clear: even for $Pb+Pb$ collisions at
CERN SPS  $\sqrt{s}/A = 17$ GeV,  the perturbative QCD parton production
is in our model anything but negligible, contrary to wide-spread belief. In fact, it
is responsible for about half of the final particle yield in the mid-rapidity region.
\medskip

Fig. 5 elucidates the second part of the above questions,
by detailing the time development of the particle production throughout
the collision history. Shown are the time-dependent total numbers
of materialized plus produced partons $N_{parton}$,
of pre-hadronic clusters $N_{cluster}$ from parton-hadron conversion,
and of produced hadrons $N_{hadron}$.
The time $t=0$ corresponds to the point of nuclear contact. 
Shortly thereafter parton materialization
by hard scatterings
produces a burst of partonic excitations $N_{parton}$ up to
$t  \, \lower3pt\hbox{$\buildrel <\over\sim$}\,0.1$ $fm$.
This initial production is then further 
enhanced by subsequent cascading and gluon emission,
which reaches a maximum around  $t \approx 1$ $fm$.
It follows then a decrease of the number of 
partons due to  formation of pre-hadronic clusters via parton coalescence,
thus feeding the buildup of $N_{cluster}$. 
The kink around $t\approx 3$ $fm$ marks the setting in of
soft cluster production from non-materialized initial-state partons
which mimics the underlying fragmentation of the nuclear remnants.
The following decays of these cluster  into primary
hadrons populate the yield $N_{hadron}$, continuing up to
$t\approx 100$ $fm$ due to the formation time of hadrons produced
in the decays.
As a consequence, 
a mixed system of clusters and already formed hadrons exists
for a long time, during which the `afterburner' cascading
of interacting clusters and hadrons is active.
The final hadron yield emerges only after about
$t\approx 200$ $fm$ when the free-streaming regime is reached.
\bigskip

\begin{minipage}[t]{17.0cm}
\begin{figure}
\epsfxsize=225pt
\centerline{ \epsfbox{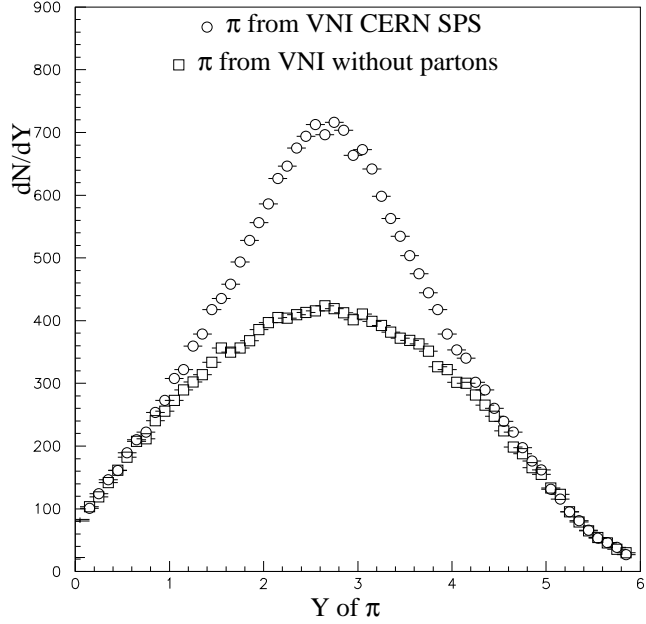} }
\vspace{-0.75cm}
\caption{
         Rapidity distribution of total pion yield in $Pb+Pb$
         collisions at CERN SPS energy $E_{cm}/A = 17$ GeV: 
         `Circles' represent the sum of contributions from
          the parton cascade and the soft 
          fragmentation of the beam remnants. `Squares' indicate
          the contribution from the beam fragmentation only.
          The difference reflects the contribution of the 
          parton cascade alone.
         \label{fig:fig4}
         }
\end{figure}
\end{minipage}
\begin{minipage}[b]{17.0cm}
\begin{figure}
\epsfxsize=265pt
\centerline{ \epsfbox{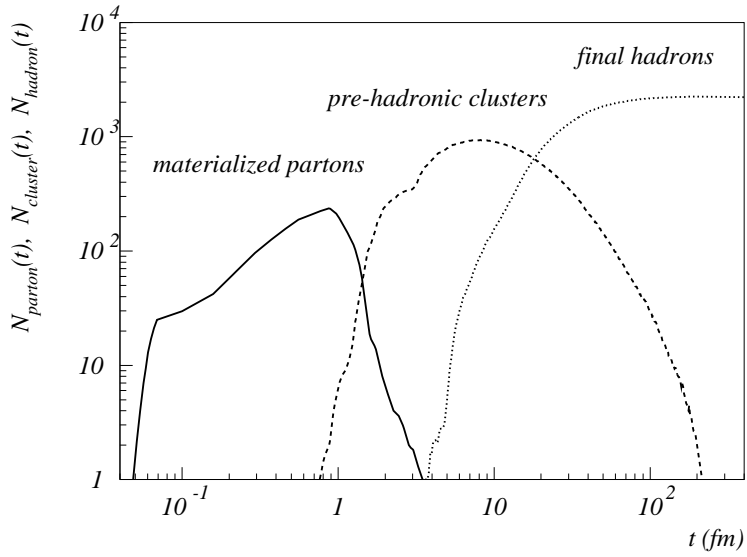} }
\vspace{-2.75cm}
\caption{
         Time evolution of the  total numbers of produced partons 
         $N_{p}$, pre-hadronic clusters $N_c$, and hadrons $N_h$ during
         $Pb+Pb$ collisions at $E_{cm}/A = 17$ GeV. The time refers
         to the center-of-mass frame of the colliding nuclei.
         \label{fig:fig5}
         }
\end{figure}
\end{minipage}

\newpage

\subsubsection{{\bf Effects of hadronic `afterburner' cascading}}
\smallskip

Next we turn to our second question of the Introduction, concerning
the effects of the `afterburner' cascade that involves re-interactions
among produced pre-hadronic clusters and already formed hadrons.
We discuss how these effects show up in  the rapidity distributions
and in the transverse mass spectra of final-state hadrons.
\medskip

\centerline{{\small{\bf a) Rapidity spectra}}}
\smallskip

Figs. 6 and 7 summarize the characteristic patterns
of hadron production with and without `afterburner' cascading of 
pre-hadronic clusters and hadrons after the parton cascade stage.
Shown are the rapidity distributions of different meson and baryon species
from $Pb+Pb$ collisions at CERN SPS energy.
Throughout, we represent by `circles' (`squares') the results from simulations
without (with) `afterburner' cascading. The solid-line histograms are results from VENUS,
which we take as representative for the CERN SPS data (the 
corresponding real data
were mostly not available for publication at this point).

Two remarkable features of the `afterburner' cascading 
are inherent in the rapidity spectra, namely,
the effect of pion and kaon absorption and associated resonance production,
and, the effect of double, or even multiple, scattering of produced
hadrons and their excitations.
Firstly, the effect of $\pi$ and $K$ absorption in $\pi(K)+cluster$ or
$\pi(K)+nucleon$ collisions is to transfer transverse energy from $\pi$ or $K$ 
via 
$\pi+C\rightarrow C^\ast\rightarrow X$ or $\pi+N\rightarrow N^\ast\rightarrow X$,
$K+N\rightarrow S +X$, where $N^\ast$ is a non-strange resonance such as $\Delta$,
and $S$ a strange baryon, such as $\Lambda,\Sigma, \Xi$.
These `chemical' reactions in effect also decrease the primary $\pi$, $K$ yield
by feeding into the non-strange and strange baryon sector, respectively,
as is evident from the  spectra in Fig. 6 and 7.
Secondly, the effect of rescatterings among clusters and hadrons
generate additional deflection and slowing down of the particles, so that the
final particle yields are concentrated more at central rapidities, as it would be the
case without re-interactions.
Both effects also give naturally rise to the phenomenon of `baryon stopping', as discussed
in the Introduction, since secondary interactions  involving
protons and neutrons from the
nuclei shift those baryons towards mid-rapidity.
\smallskip

In Fig. 6, the pion distribution clearly exhibits the effect of absorption
through `afterburner' cascading by the reduction 
of the primary pion yield before(!) `afterburner' cascading.
The effect is evident throughout the entire rapidity range $0 \le y \le 6$,
but is particularly prominent around mid-rapidity where it is about 30 $\%$.
In the kaon distribution, on the other hand, this absorption feature,
although similarly present, is hardly visible,
because it is counteracted by  additional kaon production from 
meson-meson scatterings $M+M\rightarrow K+\overline{K}$ and meson-baryon collisions
$M+B\rightarrow \overline{K}+S$ and $M+S\rightarrow K+B$.
Consistent with the $\pi$ and $K$ absorption by nucleons is the loss
of protons (both initial state and newly pair-produced ones) around
$y \approx 1.5$ and $y\approx 3.5$. For the same reason, the amount
of anti-protons from pair production is substantially decreased in the
central region $2 \le y \le 4$.

In Fig.7, the patterns of $\Lambda$ and $\overline{\Lambda}$ production
is similar to the ones of $p$ and $\overline{p}$.
The simple reason is that the $\Lambda$'s can be produced
via kaon absorption by protons, or in pairs $\Lambda\overline{\Lambda}$
just as $p\overline{p}$. It is therefore no surprise that the 
$\Lambda$ and $\overline{\Lambda}$ spectra are roughly proportional
to the $p$ and $\overline{p}$ densities, respectively.
A very different behavior, however, is exibited in
$\Sigma$ and $\Xi$ production, namely, a substantial increase
through the `afterburner' cascading of the
primary yield emerging from the hadronization of the parton cascade.
This strong population of $\Sigma$'s and $\Xi$'s arises naturally
from the hadronic re-interactions and resonance production, as well
as absorptive processes.
\smallskip

Overall we may conclude (viewing the solid-line histograms in Figs. 6 and 7 as 
reference to the shape of the actually measured distributions at CERN SPS)
that the above effects show
a general tendency of the `afterburner' cascading 
to push the primary hadron spectra after the parton cascade 
towards the ones actually observed at CERN.
The agreement of the results including the `afterburner' cascade is quite decent.
\medskip

\centerline{{\small{\bf b) Transverse mass spectra}}}
\smallskip

The particle distributions in the transverse mass variable $m_\perp = \sqrt{m^2+p_\perp^2}$
provide important information on the pattern and the degree of harnessing initial longitudinal
momentum and energy  into transverse direction.
For example, in hadronic collisions, where re-interactions are negligible on
both the parton and hadron level, a typical power-law behavior
$\propto p_\perp^{-n}$ ($n = 4-6$) 
at large $p_\perp$ is observed, which becomes increasingly
prominent as the beam energy is increased.
This characteristic $p_\perp$-production pattern indicates the contribution 
from perturbative QCD parton-parton collisions with associated (mini)jet production.
On the other hand, in nucleus-nucleus collisions, the discussed effects of
re-interactions among partons as well as hadrons, lead to a dampening of
the large-$p_\perp$ tail as well as to an overall
steepening, and hence to a characteristic exponential-type $p_\perp$-distribution.
In our model, the effects  of re-interactions on $p_\perp$-production
may be exhibited by comparing nuclear 
collisions with and without `afterburner' cascading. 
\smallskip

Fig. 8 vividly reflects the intuitive expectation that rescatterings
redistribute the particle momenta such that low-$p_\perp$ hadrons
are accelerated in transverse direction by $p_\perp$ kicks towards
larger $m_\perp$ and large-$p_\perp$ particles 
in the mean loose part of their
transvere momentum through repeated collisions.
This redistribution is nothing but the tendency of the
multi-particle system to drive towards
kinetic equilibrium through cascading.
The `equilibration effect', which twists the $m_\perp$-slopes,
is visible in all three spectra, $\pi$, $K$, and $p$,
most prominently however in the proton spectrum.
The reason is that the initial-state protons carry large fractions
of the longitudinal beam momentum, so that any scattering will 
cause a significant deflection with corresponding effect on the
$m_\perp$-distibution. Pions and kaons on the other hand, 
are produced as secondaries mostly isotropic around
mid-rapidity, and hence additional $p_\perp$ kicks 
wash away in the statistical mean, so that the effect on the
$m_\perp$-spectra is small as compared to the case of protons.
\smallskip

Again, we may conclude from Fig. 8, that the inclusion of `afterburner' 
cascading gives a fairly good agreement with the 
data (solid-lines), whereas the neglect of final-state interactions
in the case of protons is clearly off the data.


\begin{minipage}[t]{8.75cm}
\begin{figure}
\epsfxsize=225pt
\centerline{ \epsfbox{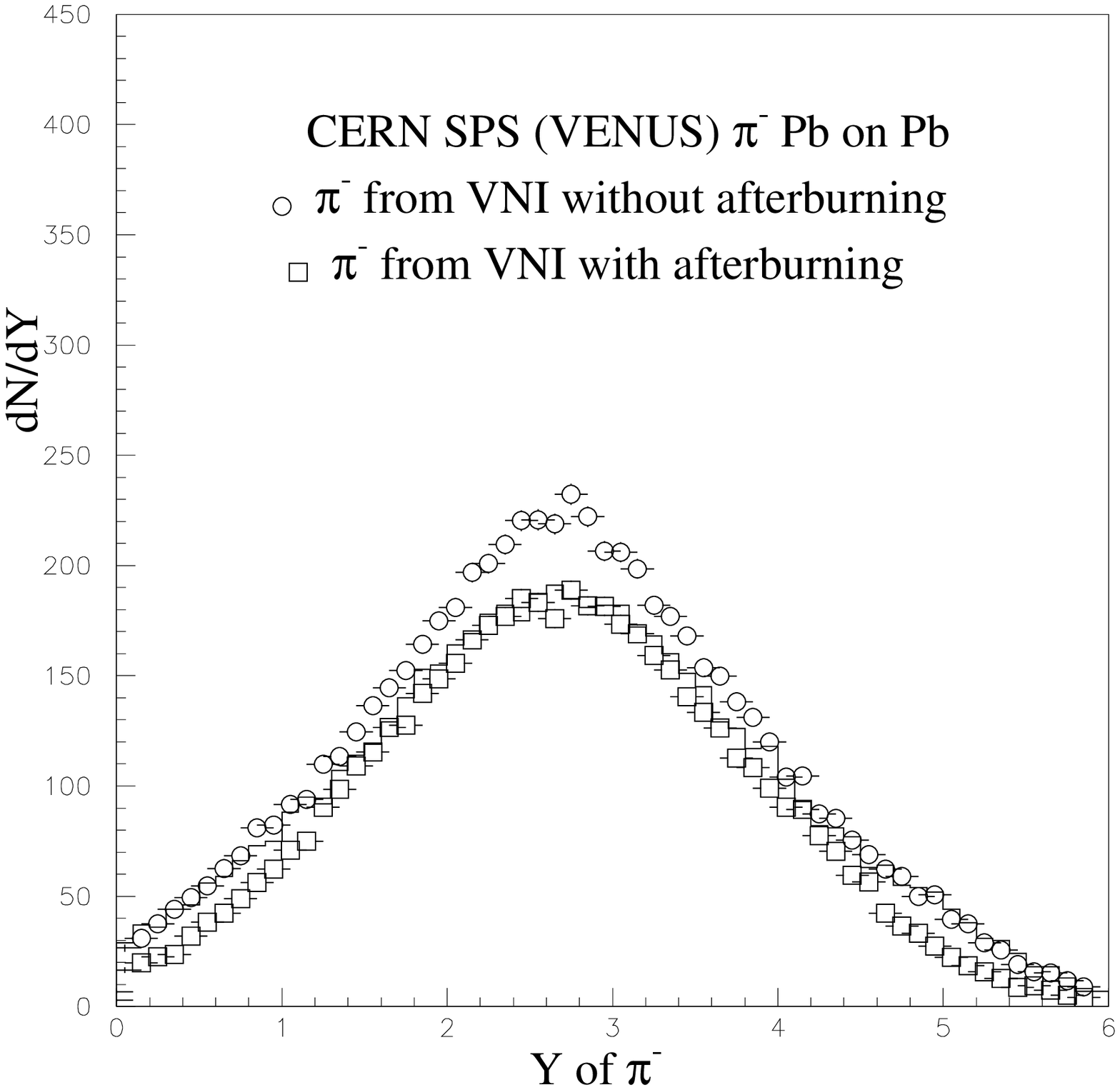} }
\end{figure}
\end{minipage}
\begin{minipage}[t]{8.75cm}
\begin{figure}
\epsfxsize=225pt
\centerline{ \epsfbox{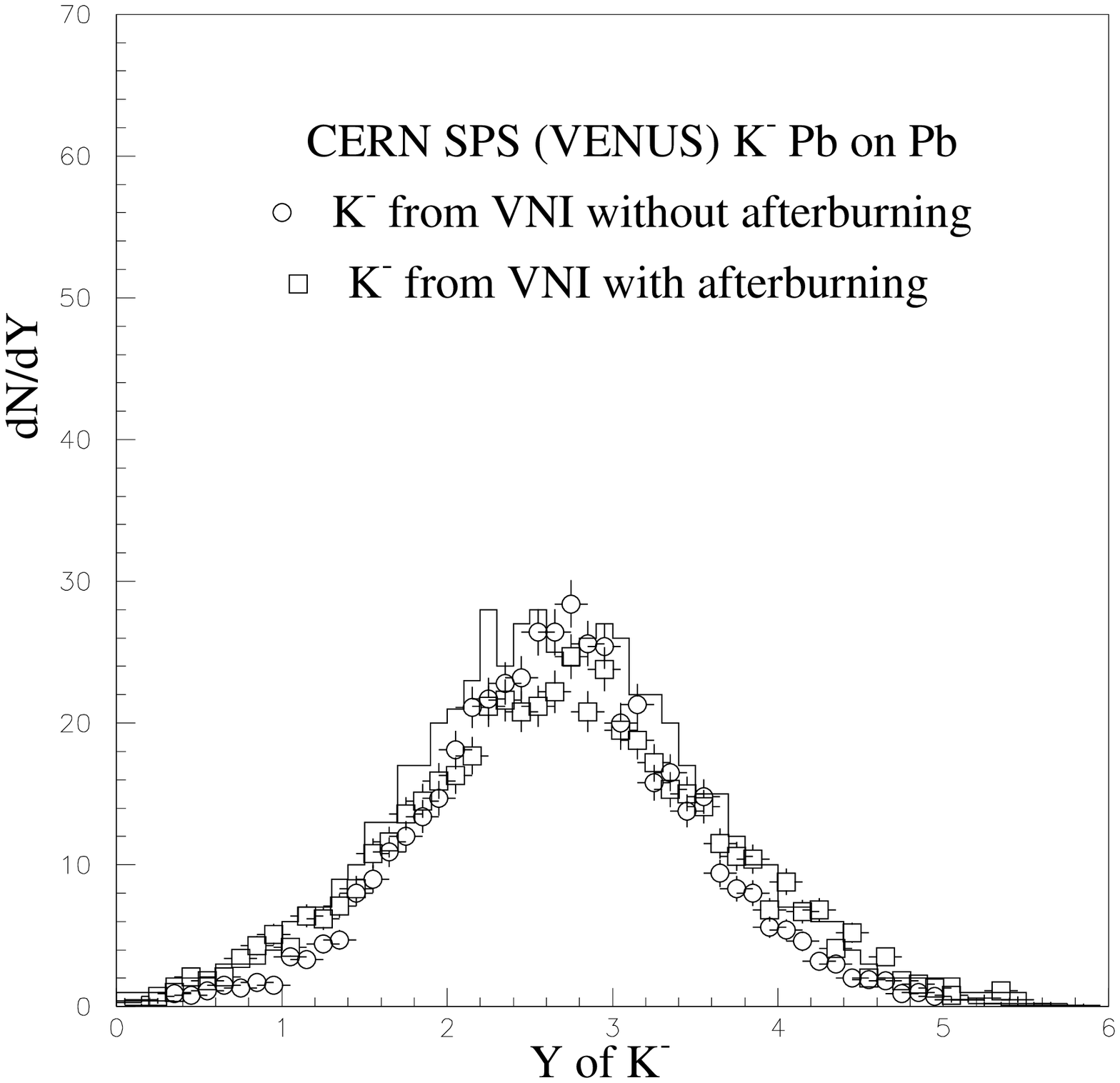} }
\end{figure}
\end{minipage}

\begin{minipage}[b]{8.75cm}
\begin{figure}
\epsfxsize=205pt
\centerline{ \epsfbox{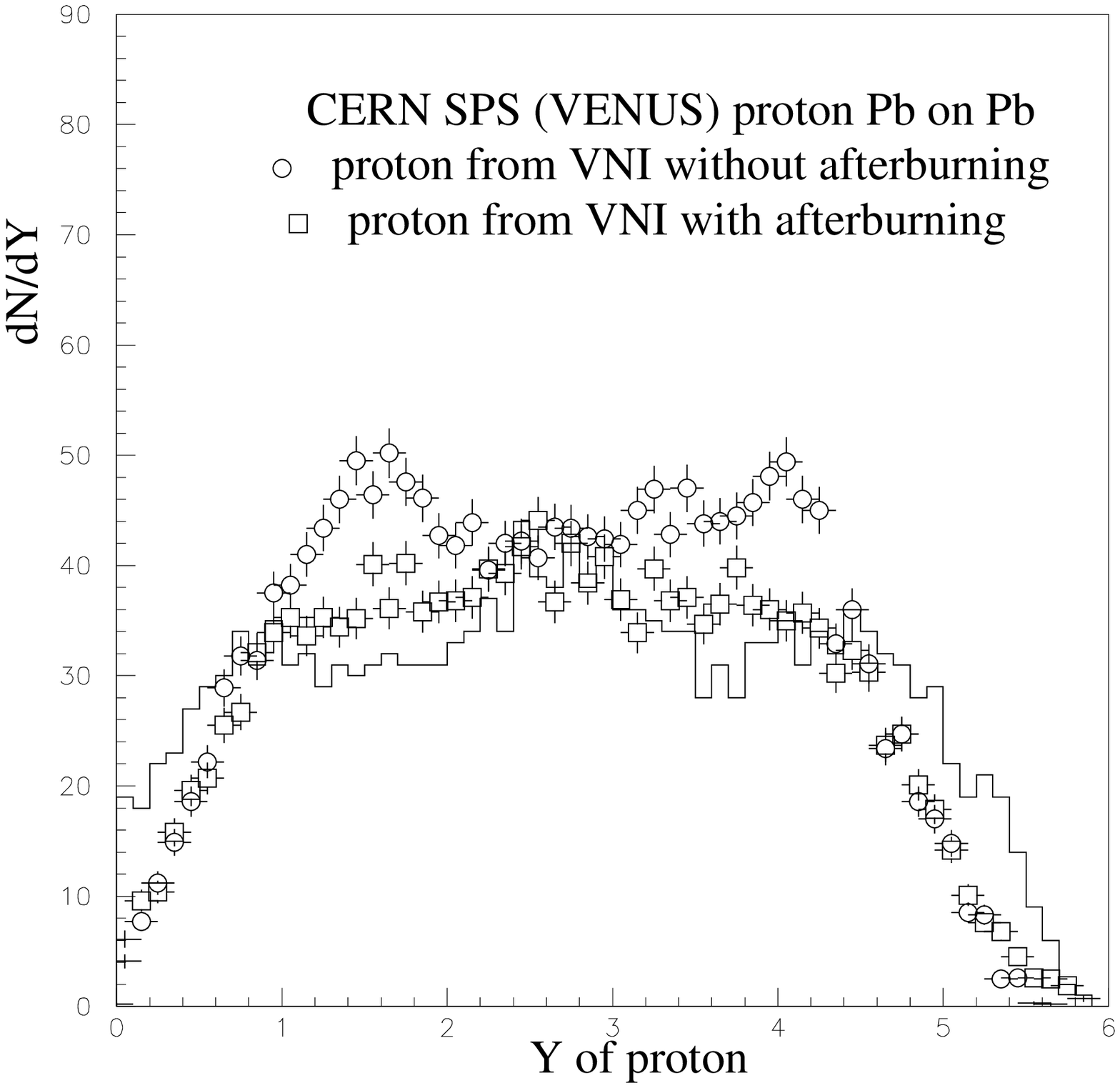} }
\end{figure}
\end{minipage}
\begin{minipage}[b]{8.75cm}
\begin{figure}
\epsfxsize=205pt
\centerline{ \epsfbox{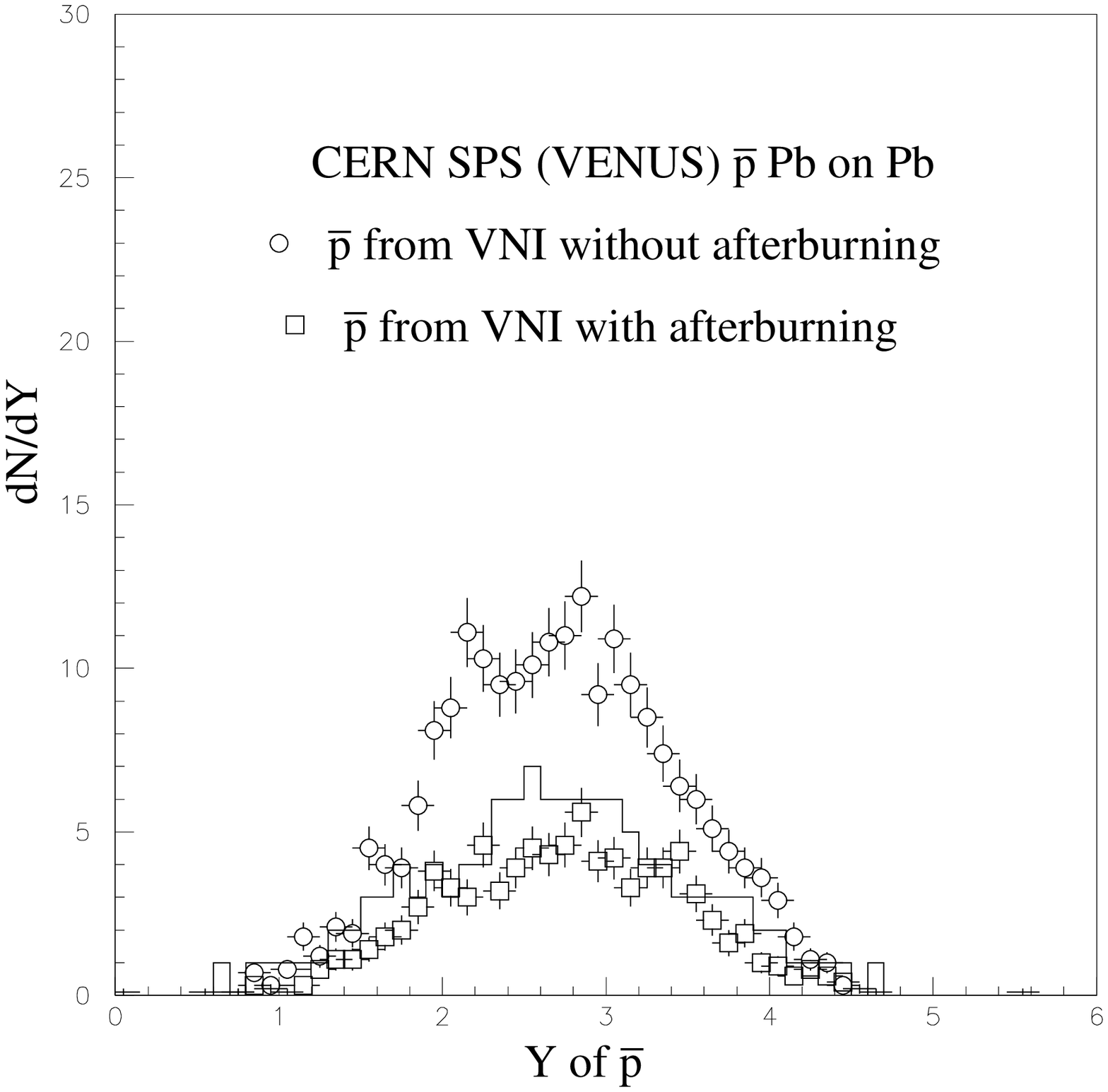} }
\end{figure}
\end{minipage}

\setcounter{figure}{5}
\begin{figure}
\vspace{-0.75cm}
\caption{
         Rapidity spectra for $\pi^-$, $K^-$, $p$ and $\bar{p}$
         without (circles) and with (squares) `afterburner' cascading
         in $Pb+Pb$ collisions at CERN SPS energy.
         \label{fig:fig6}
         }
\end{figure}

\newpage

\begin{minipage}[t]{8.75cm}
\begin{figure}
\epsfxsize=225pt
\centerline{ \epsfbox{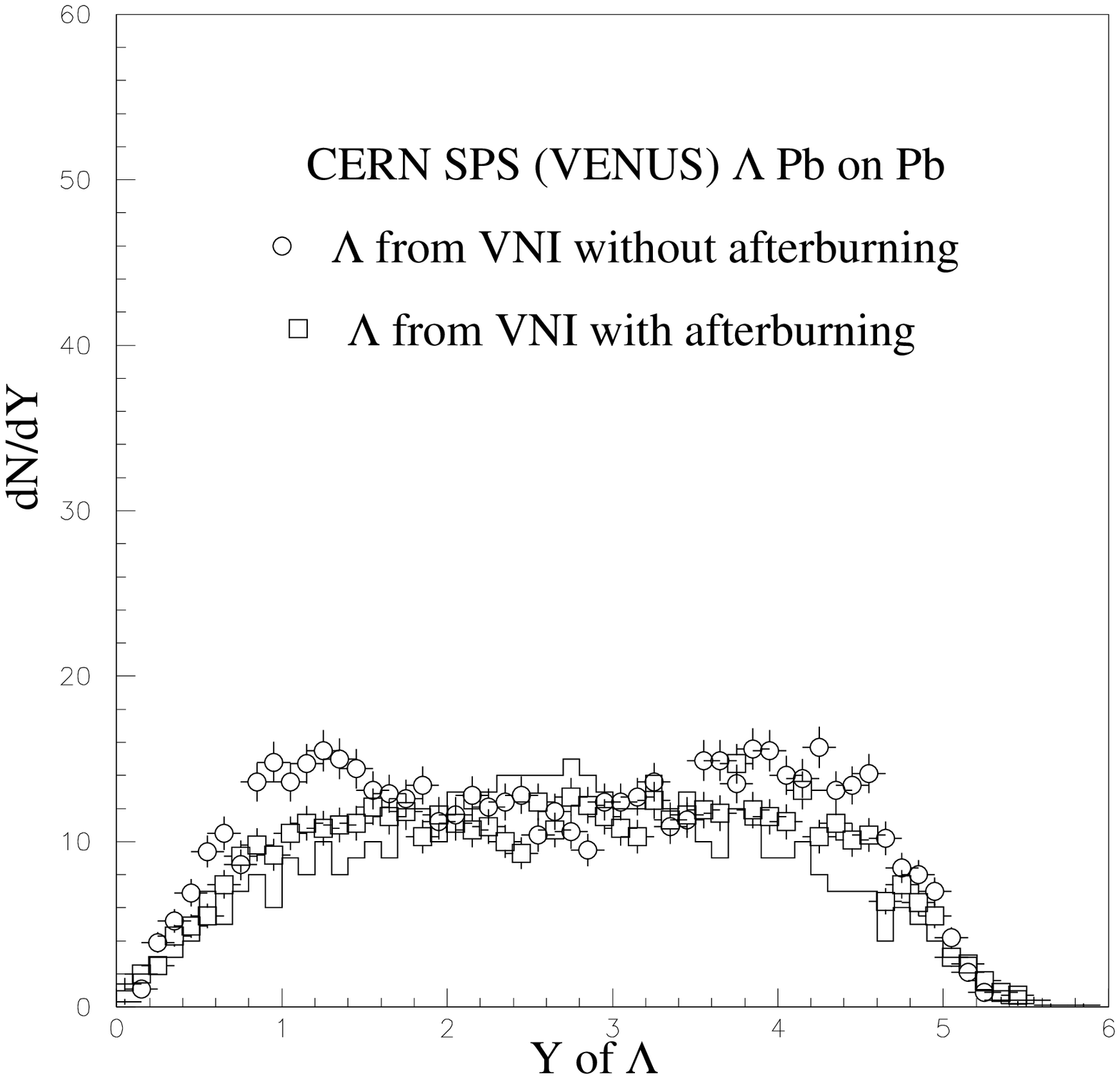} }
a
\end{figure}
\end{minipage}
\begin{minipage}[t]{8.75cm}
\begin{figure}
\epsfxsize=225pt
\centerline{ \epsfbox{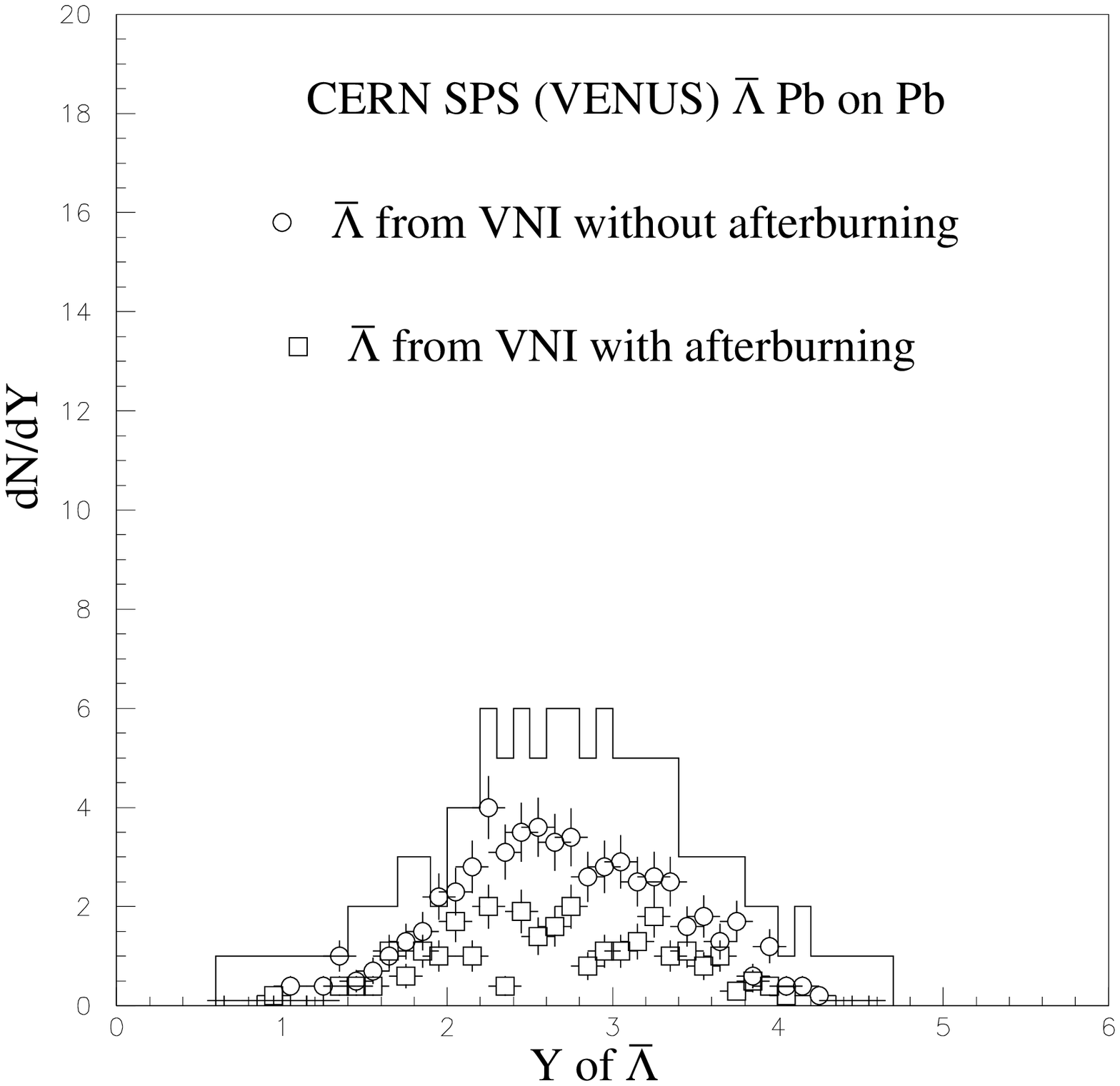} }
\end{figure}
\end{minipage}

\begin{minipage}[b]{8.75cm}
\begin{figure}
\epsfxsize=225pt
\centerline{ \epsfbox{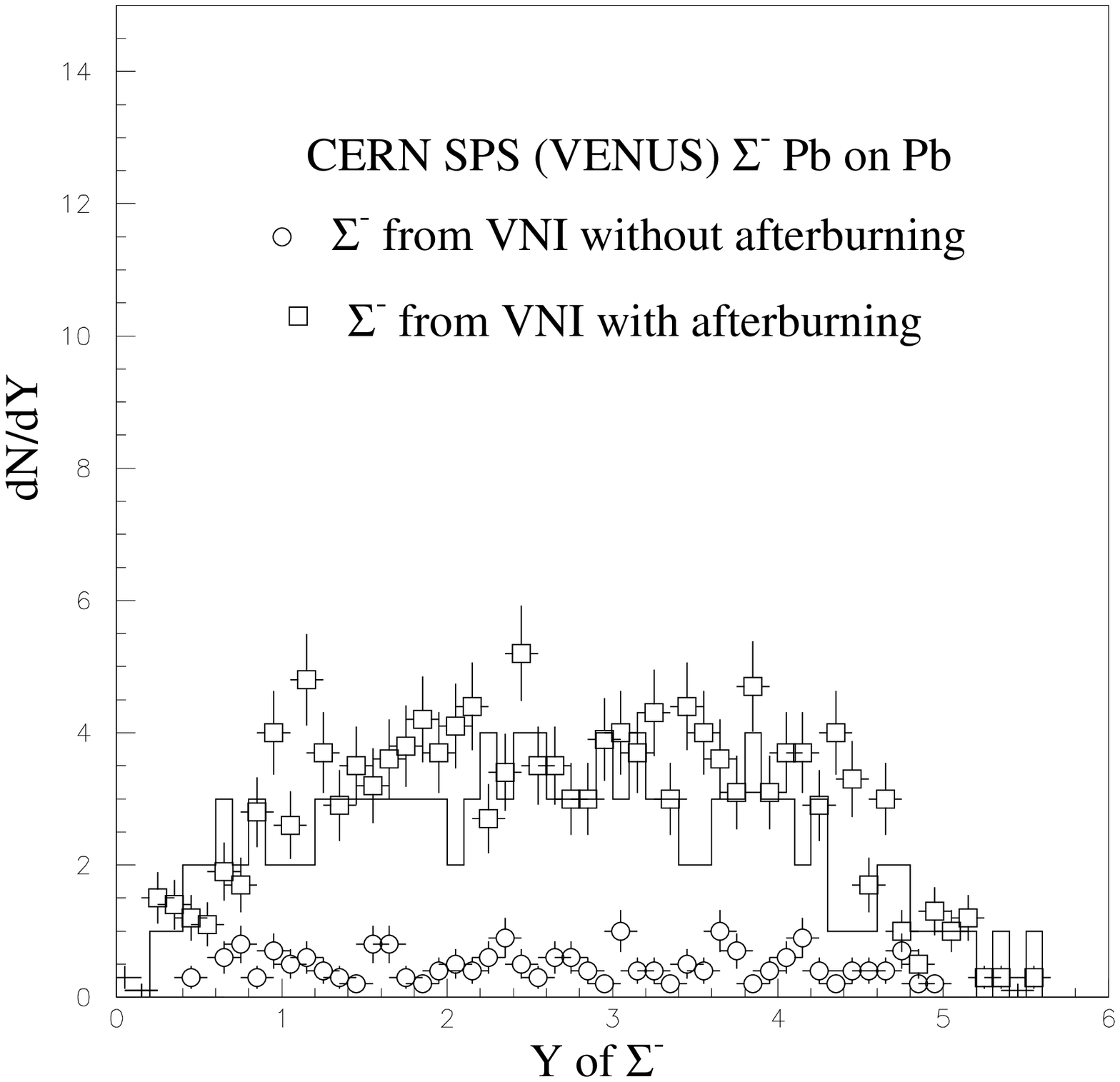} }
\end{figure}
\end{minipage}
\begin{minipage}[b]{8.75cm}
\begin{figure}
\epsfxsize=225pt
\centerline{ \epsfbox{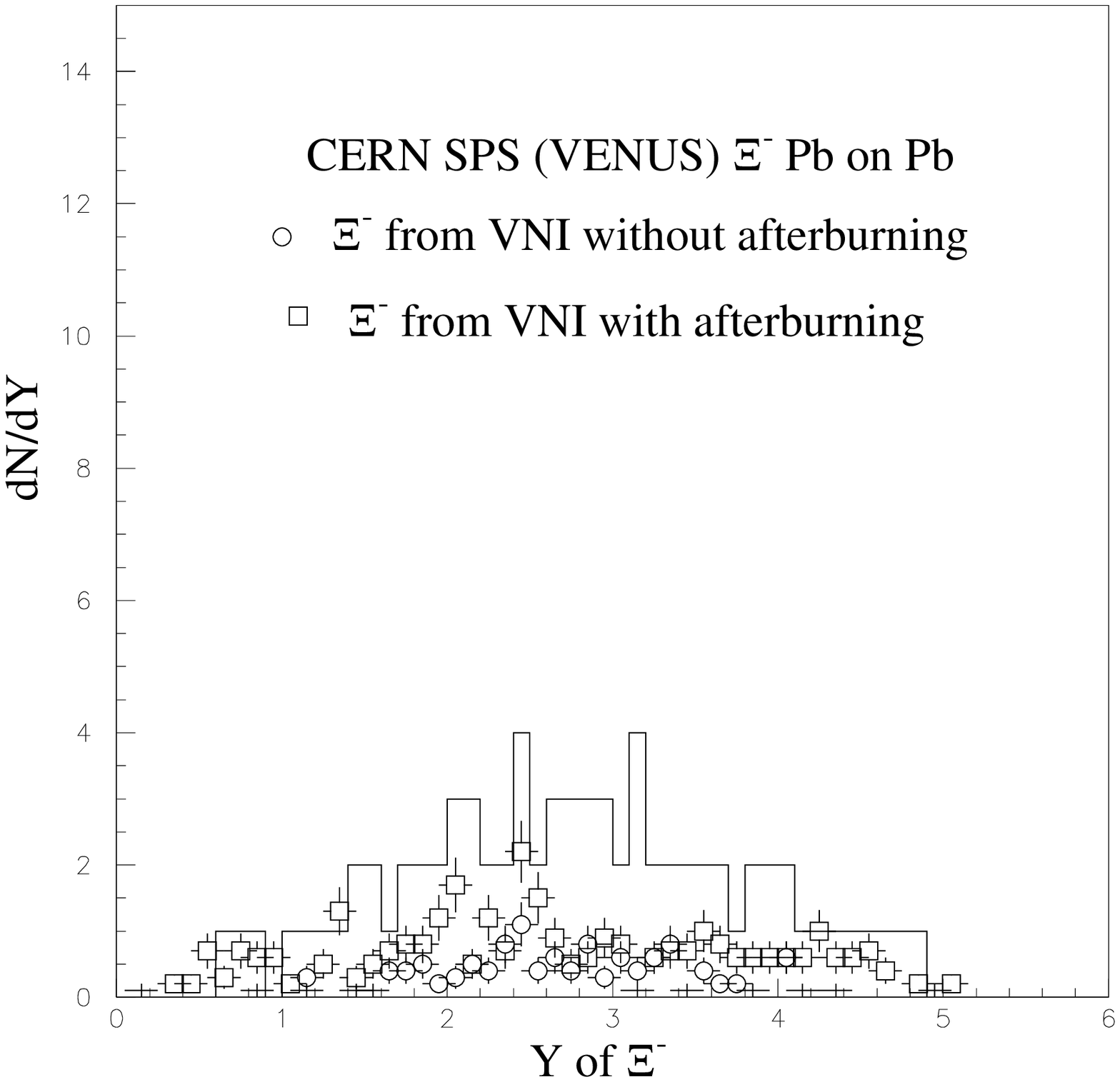} }
\end{figure}
\end{minipage}

\setcounter{figure}{6}
\begin{figure}
\vspace{-0.75cm}
\caption{
         Rapidity spectra for $\Lambda$, $\bar{\Lambda}$,
         $\Sigma^-$, $\Xi^-$
         without and with afterburner' cascading
         in $Pb+Pb$ collisions at CERN SPS energy.
         \label{fig:fig7}
         }
\end{figure}

\newpage

\begin{minipage}[t]{8.75cm}
\begin{figure}
\epsfxsize=225pt
\centerline{ \epsfbox{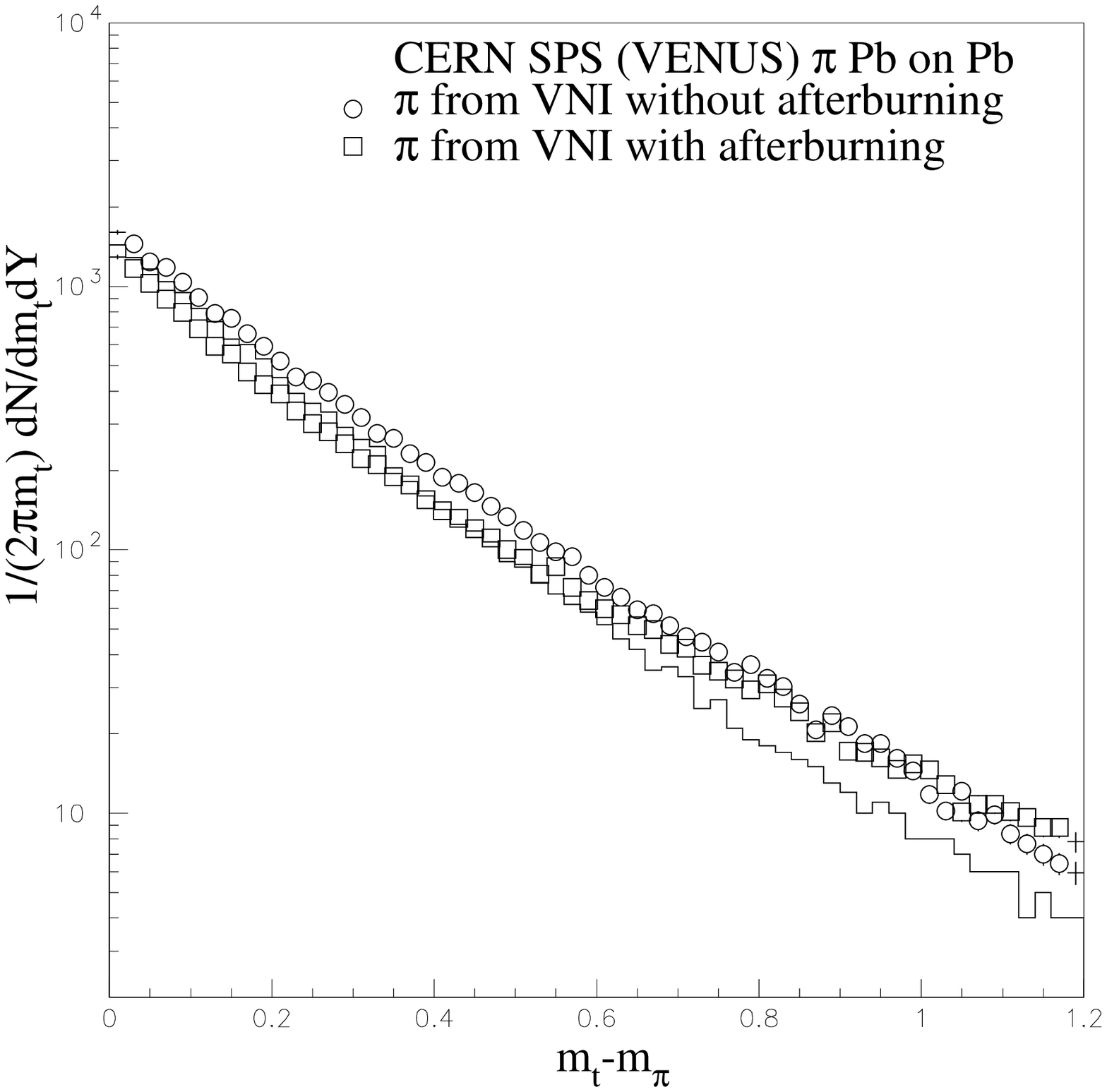} }
a
\end{figure}
\end{minipage}
\begin{minipage}[t]{8.75cm}
\begin{figure}
\epsfxsize=225pt
\centerline{ \epsfbox{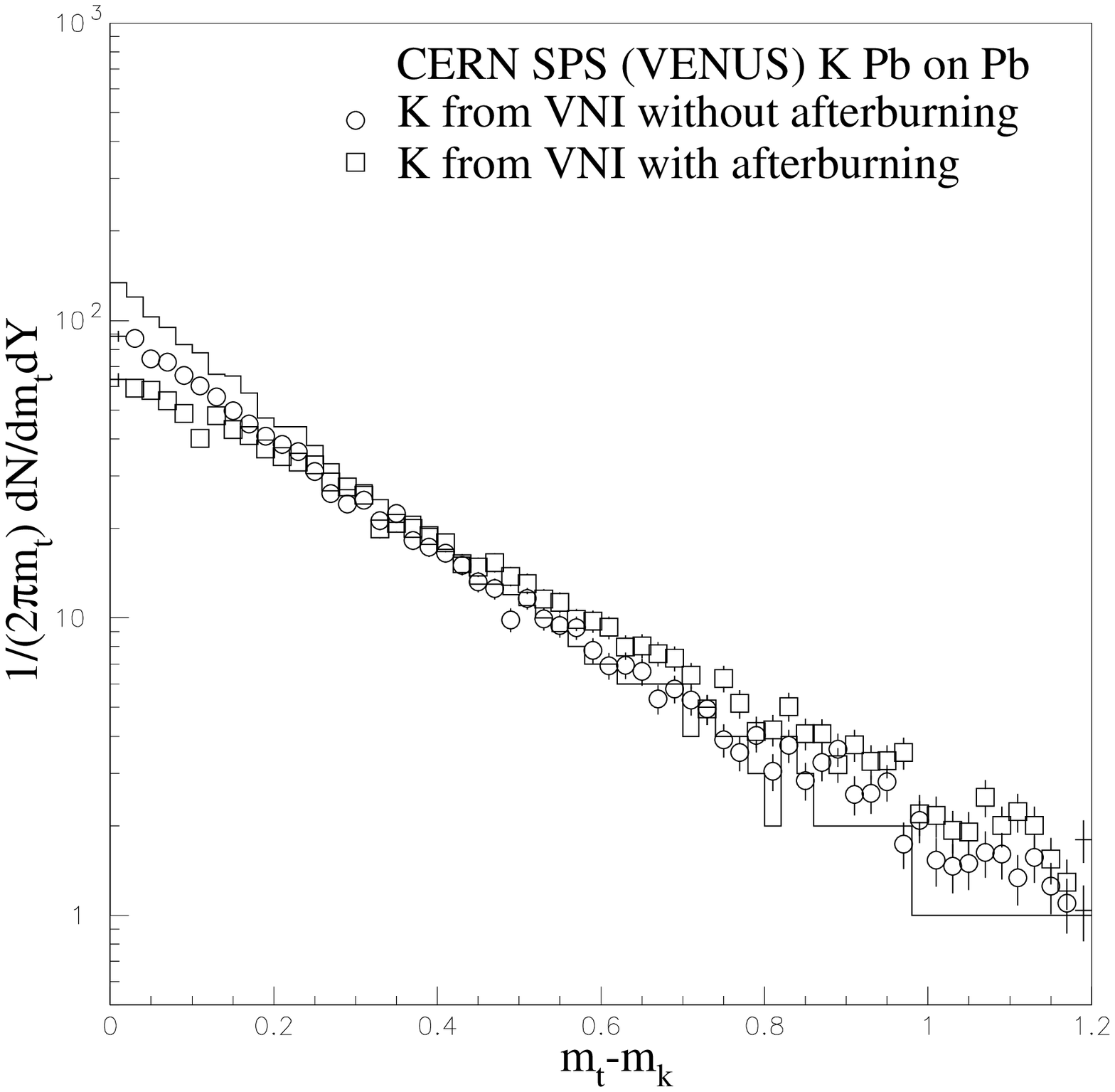} }
\end{figure}
\end{minipage}

\begin{minipage}[b]{8.75cm}
\begin{figure}
\epsfxsize=225pt
\centerline{ \epsfbox{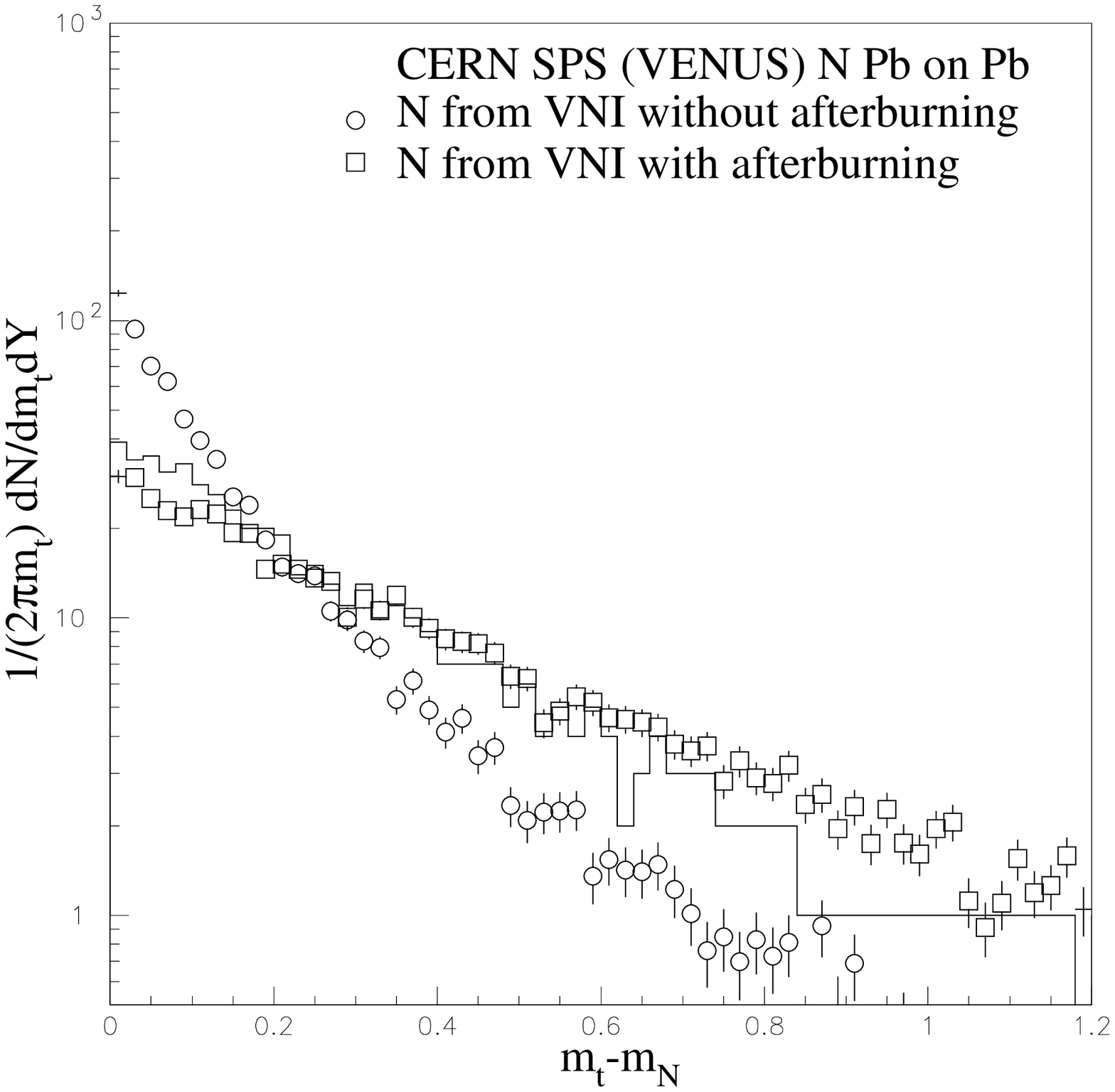} }
\end{figure}
\end{minipage}
\begin{minipage}[b]{8.75cm}
\begin{figure}
\epsfxsize=225pt
\centerline{  }
\end{figure}
\end{minipage}

\setcounter{figure}{7}
\begin{figure}
\vspace{-0.75cm}
\caption{
         Transverse mass spectra for $\pi\equiv\pi^++p^-+\pi^0$,
         $K\equiv K^++K^-$, and $N\equiv p+n$, without and with 
         afterburner' cascading in $Pb+Pb$ collisions at CERN SPS energy.
         \label{fig:fig8}
         }
\end{figure}

\newpage

\subsection{RHIC: central $Au+Au$ collisions at $E_{cm}=200$ A GeV}
\smallskip

After having gained insight within our model on the overall 
particle production features at the CERN SPS, we
now address 
$Au+Au$ collisions at more than an order of magnitude larger
collision energy
$\sqrt{s}/A = 200$ GeV.
\medskip

\subsubsection{{\bf Significance of parton production}}
\smallskip

As in our previous discussion in Sec. IIIB, 
Fig. 9 illustrates the rapidity distributon  of 
total pion production (`circles'),
resulting
from the truly perturbative QCD component of parton materializaton
and associated cascading, and from the underlying
non-perturbative component due to soft fragmentation 
of the nuclear beam remnants that consist of the
non-materialized initial state partons (`squares').
Confronting this plot for RHIC energy with the analogous Fig. 5
for CERN SPS energy, one observes that the perturbative parton
component becomes the dominant source  for
particle production at central rapidities,
yielding about 70-80 \% of the final hadrons in $|y|\le 1$.

Another important feature is the shape of the $y$-distribution
at RHIC, which is considerably stronger peaked than at CERN SPS,
due to the increased production of low-energy gluons in the former case,
both by the stronger gluon presence in the inital nuclei at
the smaller Bjorken-$x$ values probed, as well as 
by  more copious gluon emision during the parton cascading
due to larger momentum transfers of parton collisions.

Moreover,
in effect, there is no plateau-like
structure visible at mid-rapidity, even if $|y|\le 1$.
This is in strict contrast to the wide-spread
hypothesis that an approximate boost-invariant Bjorken picture
\cite{bj83}
with uniform rapidity plateau at central reapidities should emerge.
Because of the peak structure of the rapidity distribution 
the density $dN/dy|_{y=0}$
is substantially larger than commonly assumed, 
implying a very high particle density at $y\approx 0$.
However, it drops off
rapidly, so that the total, $y$-integrated yield is much less
dramatic.
\medskip

\setcounter{figure}{8}
\begin{figure}
\epsfxsize=225pt
\centerline{ \epsfbox{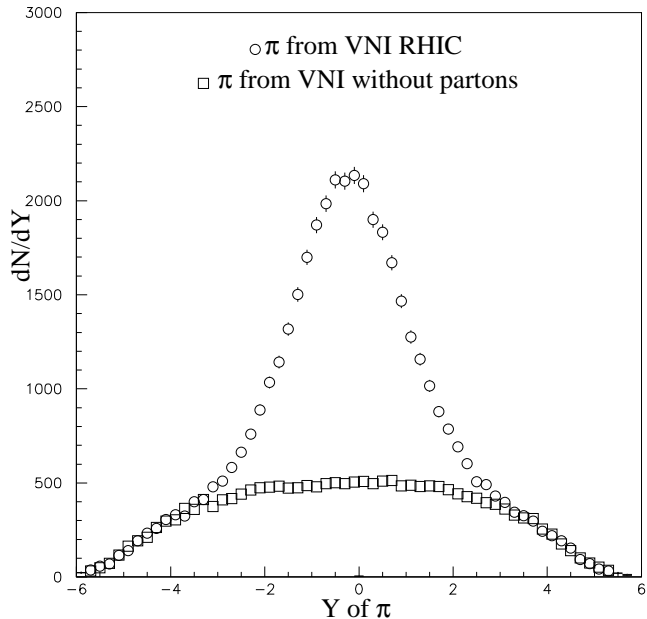} }
\vspace{-0.75cm}
\setcounter{figure}{8}
\caption{
         Rapidity distribution of total pion yield in $Pb+Pb$
         collisions at CERN SPS energy $E_{cm}/A = 17$ GeV: 
         `Circles' represent the sum of contributions from
          the parton cascade and the soft 
          fragmentation of the beam remnants. `Squares' indicate
          The contribution from the beam fragmentation only.
          The difference reflects the contribution of the 
          parton cascade alone.
         \label{fig:fig9}
         }
\end{figure}

\medskip

\subsubsection{{\bf Effects of hadronic `afterburner' cascading}}
\smallskip

Figs. 10 and 11  exhibit the effects of the `afterburner cascading'
at RHIC,  including
as before the re-interactions among prehadronic clusters and hadrons,
in correspondence to Figs. 6 and 7 for CERN SPS.
\smallskip

Fig. 10, which
shows the rapidity distributions of final-state pions, kaons and protons,
indicates that the `afterburner' cascading is equally important at RHIC
as it is at CERN SPS, neither is it less, nor seems there to be significant
increase of hadronic cascading at RHIC as one might have expected
due to  a factor of 10 larger energy per nucleon available. 
However, in view of the previous Fig. 9,
this is understandable, because a larger amount of the initial
total collision energy is harnessed during the parton cascade stage.
If that holds true at even larger energies, this would imply
a certain universal, energy-independent character of the `afterburner'
cascading.
\smallskip

The general patterns of the `afterburner' cascading effects 
at RHIC in Fig. 10 and 11
resembles the ones of Fig. 6 and 7 for CERN SPS, however, 
there is an important difference,
namely the fact that at RHIC the nucleons, mostly originating from
the nuclear beams, are pre-dominantly appearing in the  rapidity
intervals $-4 \le y\le -2$ and $2 \le y\le 4$, as is evident
from the plot of the proton distribution in Fig. 11.
Hence, the separation between the baryon-rich regions in the forward
and backward hemispheres is 
twice as large as at the CERN SPS (c.f. Fig. 6), where they peak within
$1 \le y\le 2$ and $4 \le y\le 5$,
corresponding to center-of-mass rapidities $-2 \le y\le -1$ and $1 \le y\le 2$.
As a consequence of the large density of primary nucleons
at RHIC in the rapidity intervals $-4 \le y\le -2$ and $2 \le y\le 4$,
an enhanced production of pions and kaons is generated
by re-interactions of primary nucleons:  
The $\pi$ and $K$ spectra in Fig. 10 
with `afterburner' cascading (`squares') exhibit shoulders
in these baryon-rich rapidity intervals,
as compared to the case without (`circles').
The opposite is true for the proton distibution in Fig. 11, which
becomes substantially diminished in these $y$-regions
after inclusion of the `afterburner' cascading, because  secondary 
collisions involving nucleons, 
such as  $N+\pi\rightarrow N* +\pi \rightarrow n\, \pi +X$
eat away protons and neutrons, feeding the pion yield, for instance.
\smallskip

Also shown in Fig. 11 are  the $y$-spectra of
$\bar{p}$, $\Lambda$, and $\Sigma$ for RHIC.
As compared to Fig. 7, which shows  the corresponding
spectra for CERN SPS, one recognizes  again that the main activity
for  $\Lambda$ and  $\Sigma$ production is concentrated in
$-4 \le y\le -2$ and $2 \le y\le 4$, because  the 
nucleon density is the highest there so that nucleon-induced
production of these particles is enhanced.
In contrast to this, the production of
$\bar{p}$ (and similarly for $\bar{\Lambda}$, not shown here),
is concentrated in the central region $|y|\le 2$,
as one would expect, because the $\bar{p}$ (or $\bar{\Lambda}$)
are produced via pair-production, rather than by
pion or kaon absorption of high-momentum nucleons.


\begin{minipage}[t]{8.75cm}
\begin{figure}
\epsfxsize=225pt
\centerline{ \epsfbox{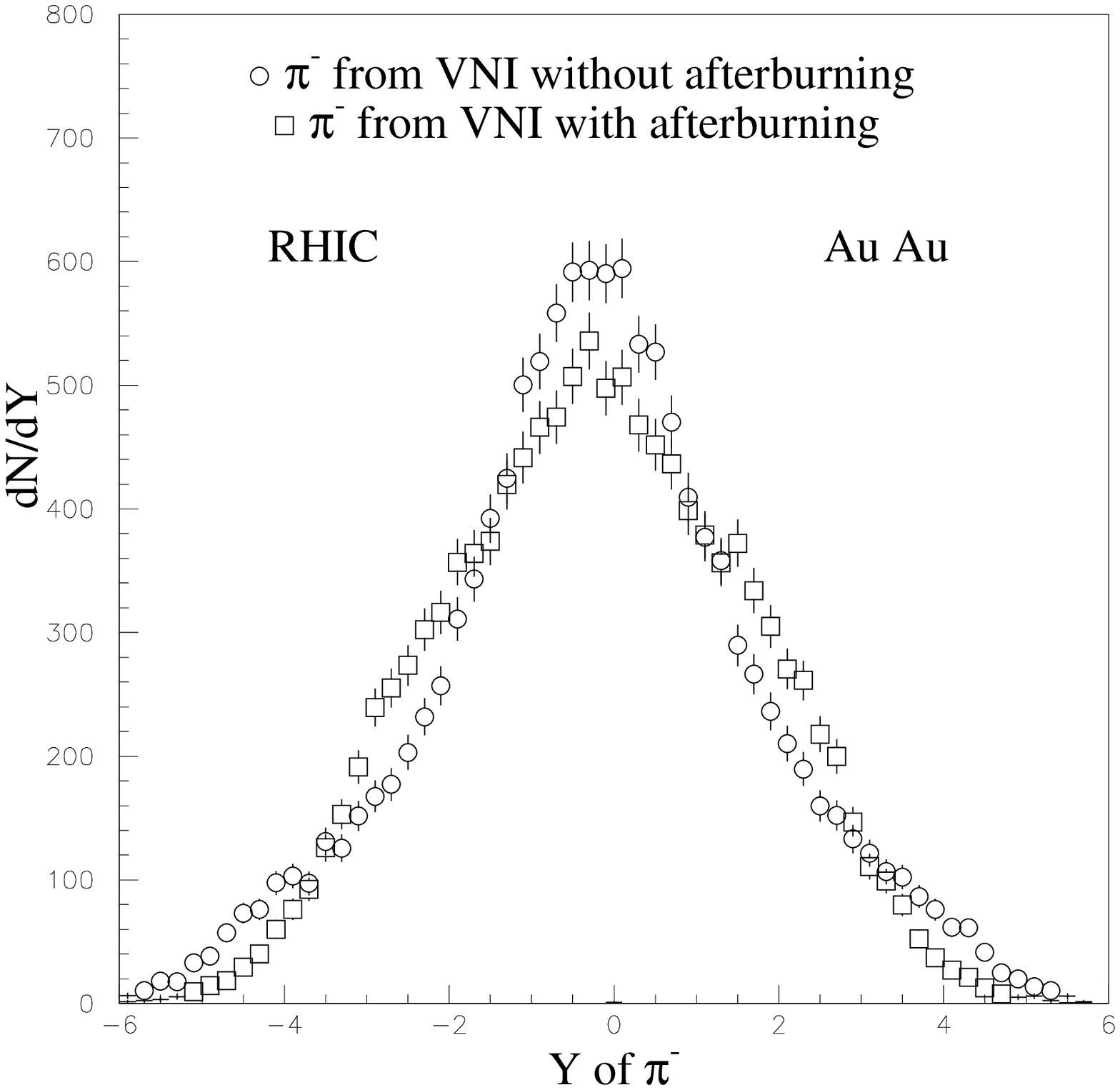} }
\end{figure}
\end{minipage}
\begin{minipage}[t]{8.75cm}
\begin{figure}
\epsfxsize=225pt
\centerline{ \epsfbox{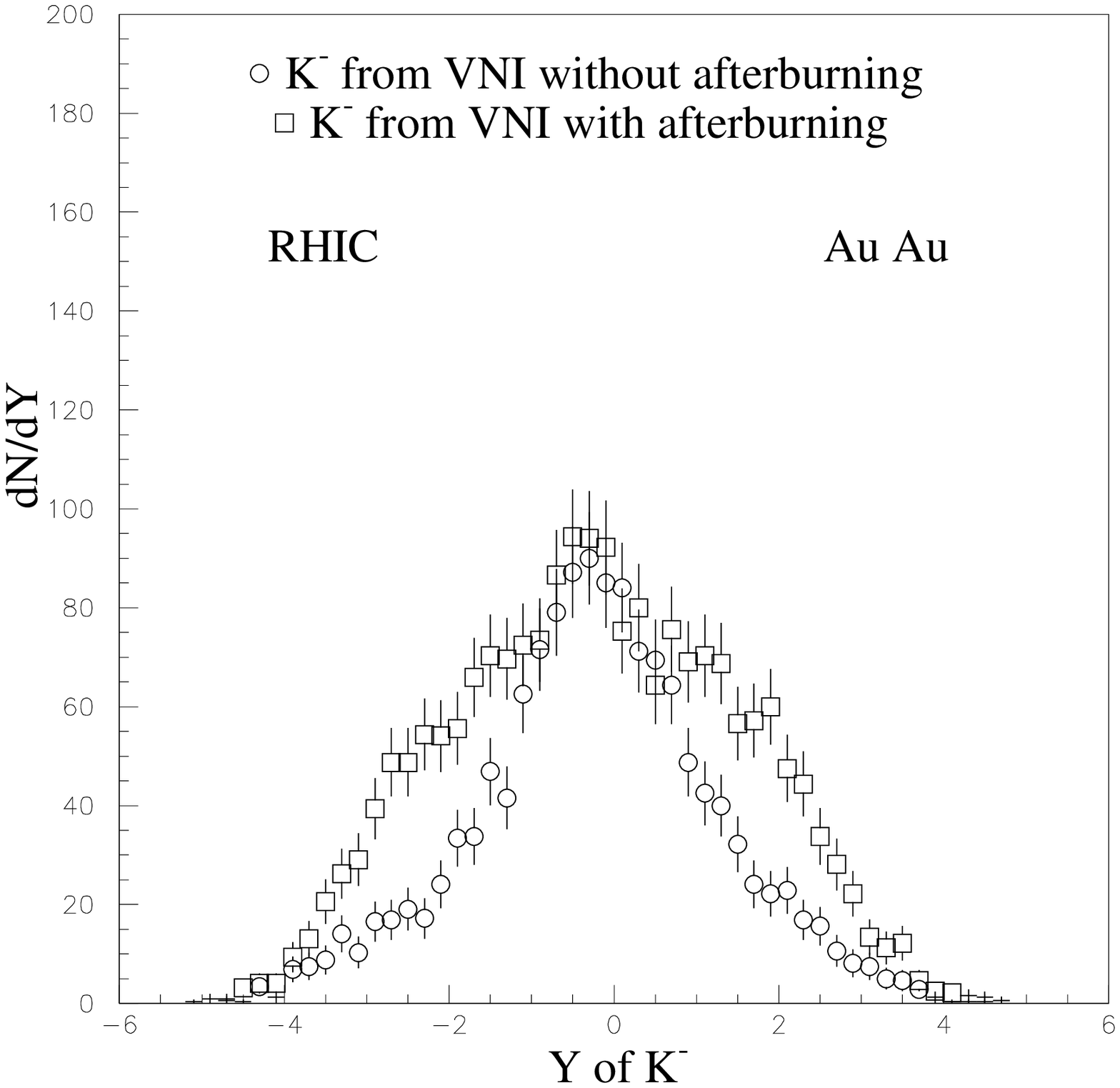} }
\end{figure}
\end{minipage}
\setcounter{figure}{9}
\begin{figure}
\vspace{-0.75cm}
\caption{
         Rapidity distributions $\pi^-$ and $K^-$ 
         without (circles) and with (squares) 
         afterburner' cascading in $Au+Au$ collisions at RHIC energy.
         \label{fig:fig10}
         }
\end{figure}

\newpage

\begin{minipage}[b]{8.75cm}
\begin{figure}
\epsfxsize=225pt
\centerline{ \epsfbox{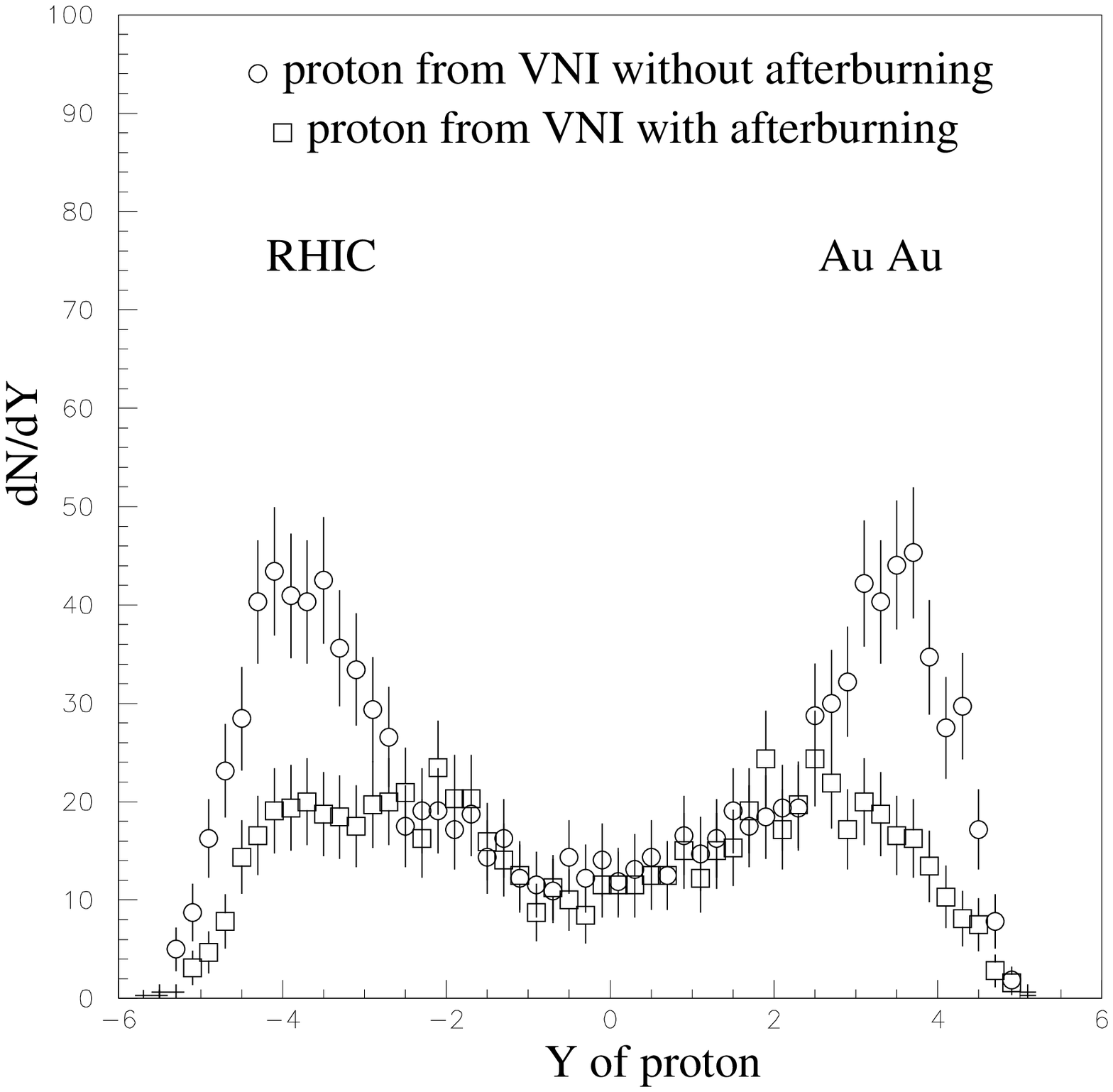} }
\end{figure}
\end{minipage}
\begin{minipage}[b]{8.75cm}
\begin{figure}
\epsfxsize=225pt
\centerline{ \epsfbox{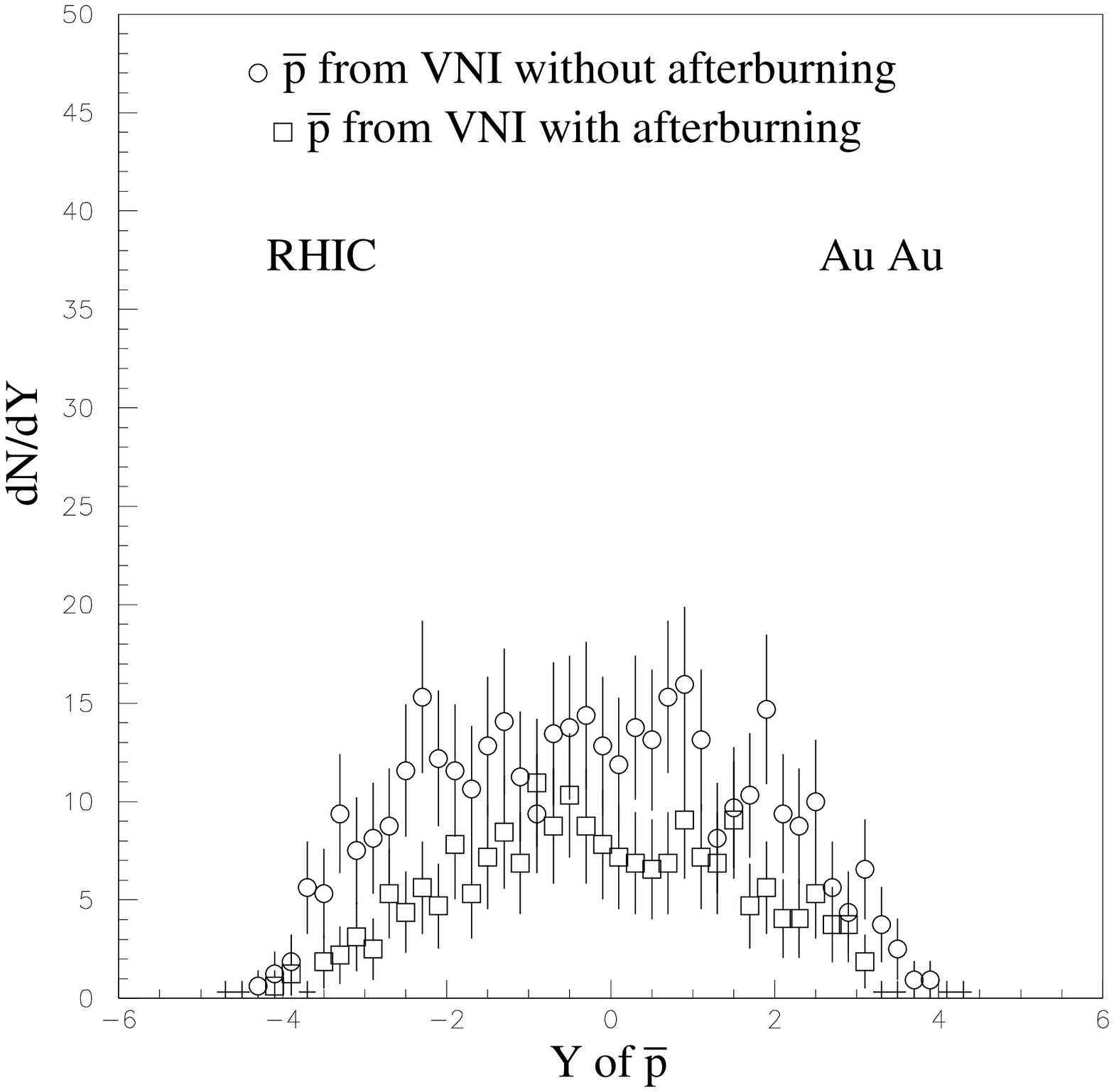} }
\end{figure}
\end{minipage}

\begin{minipage}[t]{8.75cm}
\begin{figure}
\epsfxsize=225pt
\centerline{ \epsfbox{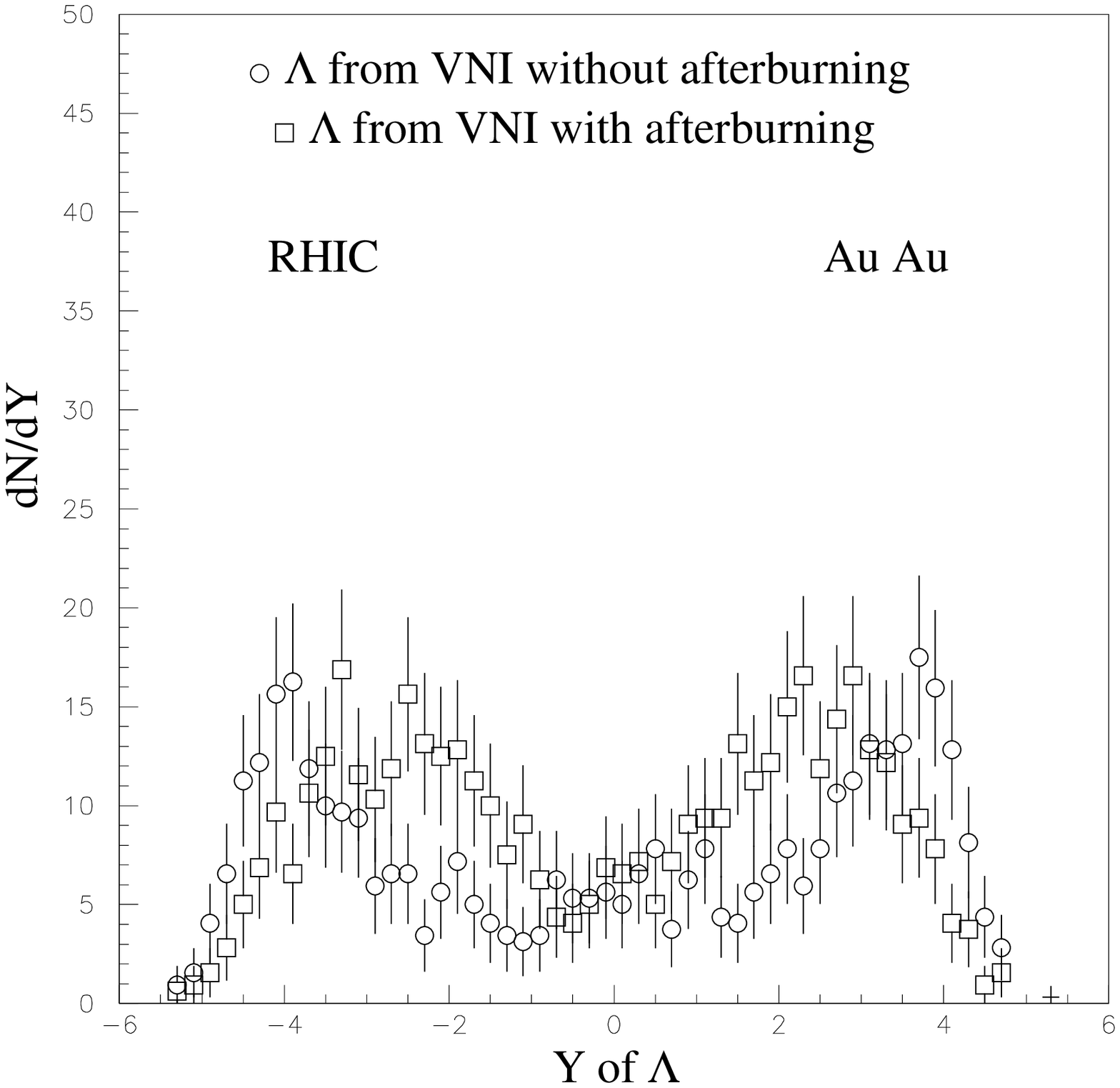} }
\end{figure}
\end{minipage}
\begin{minipage}[t]{8.75cm}
\begin{figure}
\epsfxsize=225pt
\centerline{ \epsfbox{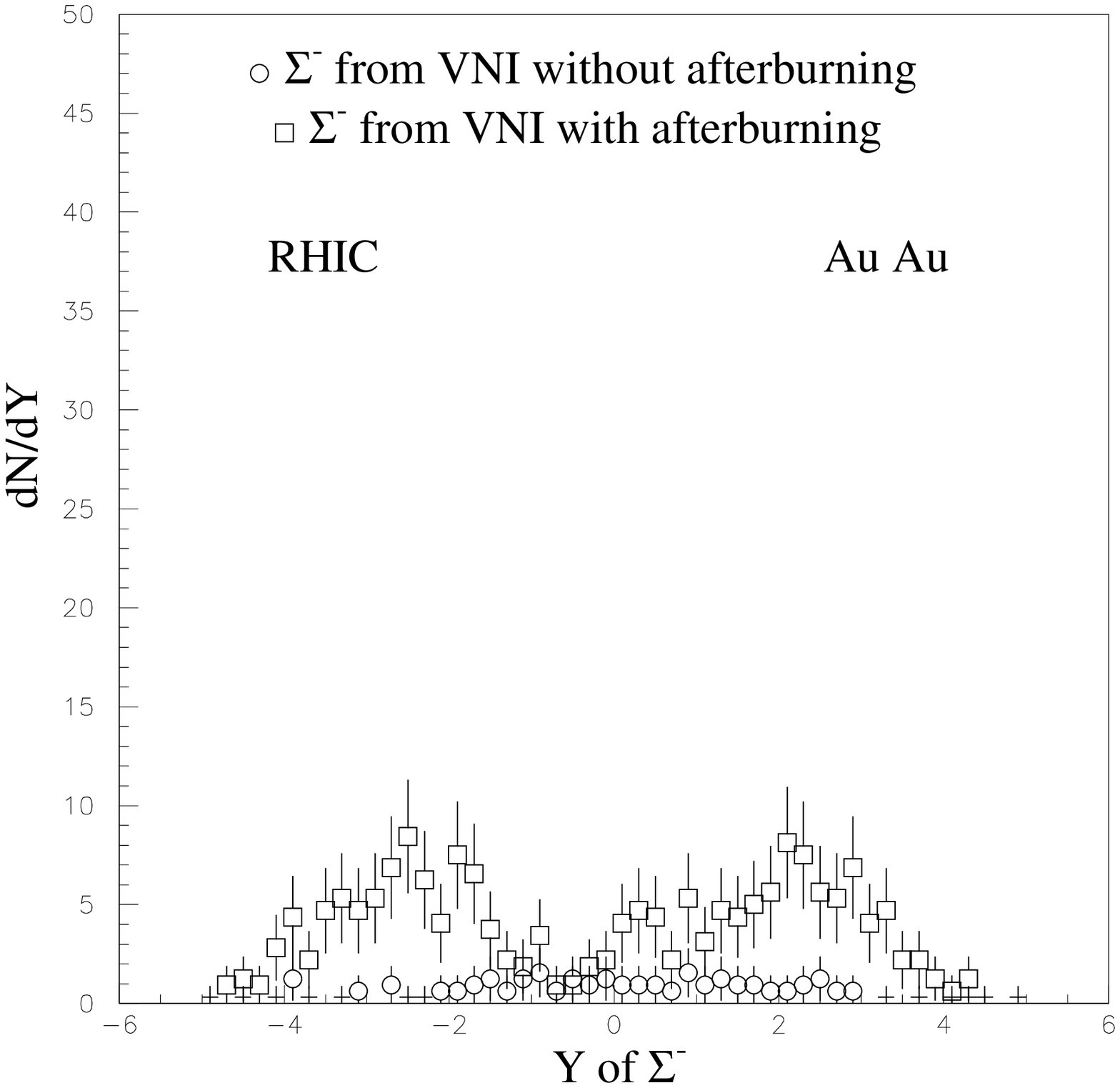} }
\end{figure}
\end{minipage}

\setcounter{figure}{10}
\begin{figure}
\vspace{-0.75cm}
\caption{
         Rapidity distributions of $p$, $\bar{p}$, $\Lambda$ and $\Sigma^-$,
         without  and with
         afterburner' cascading in $Au+Au$ collisions at RHIC energy.
         \label{fig:fig11}
         }
\end{figure}

\bigskip
\bigskip
\bigskip

\section{SUMMARY}
\smallskip

In this work we have analysed $Pb+Pb$ collisions at the
CERN SPS with beam energy per nucleon 158 GeV ($\sqrt{s}/A = 17$ GeV)
and $Au+Au$ collisions at RHIC with  collider energy $\sqrt{s}/A = 200$ GeV,
by performing Monte Carlo simulations
within the framework of a space-time model that involves
the dynamical interplay between parton production, evolution,
parton-cluster formation, and `afterburner' cascading
of formed pre-hadronic clusters plus hadron excitations.
We see this study as a first exemplary attempt to 
describe high-energy nuclear collisions on
the microscopical level of the
space-time history of parton and hadron degrees of freedom
based on QCD and supplemented by phenomenology.
We developed a `joint venture' of the Monte Carlo programs
VNI and HIJET that allowed us to trace in detail the 
time evolution 
of nuclear collisions in both position and momentum space,
from the instant of nuclear overlap to the final yield
of particles.
\smallskip

Our main conclusions may be summarized as follows:
\begin{description}
\item{(i)}
The perturbative QCD parton production and parton cascading
provides an important contribution to particle production
at central rapidities. In effect 
this partonic component almost doubles 
the amount of pions around $y=0$, as compared to 
sole contribution from the underlying soft fragmentation
of the nuclear beam particles.
\item{(ii)}
The `afterburner' cascading of pre-hadronic clusters and already formed hadrons 
which emerge from the parton cascade, as well as from
the soft fragmentation of the nuclear beam remnants,
turns out  to be most important in the
central phase-space region around mid-rapidity where the particle densities
are the largest.
The two main mechanisms of the `afterburner' cascading
are pion/kaon absorption by, and multiple
scattering of  clusters and hadrons.
\item{(iii)}
The overall agreement of our
model calculations including the `afterburner' cascading
with the observed particle spectra at the CERN SPS
is fairly good, whereas the neglect of the final-state
interactions among hadronic excitations deviates
significantly from the data. 
\item{(iv)}
In heavy-ion collisions at RHIC, we have 
absolutely no indication of a plateau-form
at mid-rapidity, not even in a narrow interval at $y=0$.
Instead the bulk of particles strongly peaks
leading to a very high particle density $dN/dy|_{y=0}$,
but rapidly dropping off for $y$ values different from zero.
\item{(v)}
In RHIC collisions, we find that
an essentially baryon-free region between
$-2 \le y\le 2$ is created, whereas the the high-momentum
nucleons appear mainly in the rapidity intervals
$-4 \le y\le -2$ and $2 \le y\le 4$ and
produce two baryon-rich peaks in the forward and backward
hemispheres, respectively.
\end{description}
Especially the latter two points would have important consequencees:
If both these effects indeed turn out  so drastic at RHIC 
as compared to  CERN SPS, then this would
imply very   favorable conditions for the formation of a 
baryon-free quark-gluon plasma in the central rapidity region,
and possibly even a baryon-rich plasma in the nucleon-dense 
fragmentation regions.
\smallskip

In view of the semi-classical particle picture
underlying  our approach, we may say that the
bulk features of particle production at CERN SPS can actually be
described without 
introducing additional prescriptions
to accomodate the observed physics
(we emphasize that we have not attempted to fine-tune our 
model to fit the data). 
On the other hand, an investigation of more
sensitive observables may well fail, in which
case we could identify this as truly new
physics beyond a simple particle-cascade description. 
In any case it would be desirable to extend and deepen this study by looking
for instance at nuclear collisions with  other beams and different energy, or
at proton-nucleus collisions. We intend to pursue this project in the
near future.

Notwithstanding, at this point we may take  the decent
description of the CERN $Pb+Pb$ data by our
mixed parton-hadron cascade as encouragement to
look forward  to RHIC or even LHC.
Since the relative importance of 
partonic and hadronic degrees are regulated by the 
multi-particle dynamics itself,
our model should provide a smooth extrapolation to
these future nuclear collisions beyond the CERN SPS.
\bigskip
\bigskip
\bigskip
\bigskip

\noindent
\centerline{\bf ACKNOWLEDGEMENTS}
\medskip

\centerline{This work was supported in part by the D.O.E under contract no.
DE-AC02-76H00016.}

\end{document}